\documentclass[12pt,a4paper]{article}
\usepackage[latin9]{inputenc}
\usepackage{times}
\usepackage{a4wide}
\usepackage{amssymb}
\usepackage{amsmath}
\usepackage{bbm}
\usepackage{dsfont} 
\usepackage{ulem} 
\usepackage{pifont} 
\usepackage{ifpdf}
\ifpdf
\usepackage[pdftex]{graphicx}
\usepackage[pdftex,unicode,implicit]{hyperref}
\hypersetup{%
  pdftitle    = {All the timelike supersymmetric solutions of all ungauged d=4  supergravities},
  pdfkeywords = {supergravity, supersymmetry, BPS},
  pdfauthor   = {P. Meessen, T. Ortin and Silvia Vaula},
  pdfcreator  = {pdf\LaTeXe\ with package \flqq hyperref\frqq},
  pdfproducer = {pdf\LaTeXe\ with package \flqq hyperref\frqq},
  pdfpagemode = None,  
  pdffitwindow= true,  
  unicode     = true,
  plainpages  = false,
  colorlinks  = true,  
  citecolor   = blue,
  urlcolor    = red,
  linkcolor   = black
}
\newcommand{\hepth}[1]{{\tt \href{http://www.arXiv.org/abs/hep-th/#1}{hep-th/#1}}}

\newcommand{\arxiv}[1]{{\tt \href{http://www.arXiv.org/abs/#1}{arXiv:#1}}}
\else
  \usepackage[dvips]{graphicx}
  \usepackage[unicode,implicit]{hyperref}
  \newcommand{\hepth}[1]{{\tt hep-th/#1}}

  \newcommand{\arxiv}[1]{{\tt arXiv:#1}}
\fi
\usepackage{tikz}
\newcommand{\FPAUO}[2]{
\tikz[scale=.13,
         Uniovi/.style={color=gray, fill=gray}
 ] {
 \fill[Uniovi] (0,0) circle (10);
 \fill[white] (0,7) circle (1.5);
 \draw[Uniovi] (-2,7.5) rectangle (2,5.5);
 \fill[white] (-0.3,6.6) rectangle (0.3,0);   
 \fill[white] ( -0.9,6.2) rectangle (.9 ,5.6);
 \fill[white] (-1.4, 5.2) rectangle (1.4, 4.6);
 \fill[white] (0,0) ellipse (3.5 and 4);
 \fill[Uniovi] (-2.5,0.3) rectangle (2.5,-0.3);
 \fill[Uniovi] (-2,2.3) rectangle (2,1.7);
 \fill[Uniovi] (-2,-2.3) rectangle (2,-1.7);
 \fill[white] (-4.5,5.5) rectangle (-2.7,4.9);
 \fill[white] (-3.9,6.1) rectangle (-3.3,4.3);
 \fill[white] (4.5,5.5) rectangle (2.7,4.9);
 \fill[white] (3.9,6.1) rectangle (3.3,4.3);
 \foreach \x in { 0,..., 3 }
   \foreach \y in { 0,...,\x}
    {
     \fill[white] (-6-\x*0.7+\y*1.4,3.5-\x *1.97) -- (-5.6-\x*0.7+\y*1.4,2.4-\x *1.97) -- (-6.4-\x*0.7+\y*1.4,2.4-\x *1.97) -- cycle;
     \fill[white] (6-\x*0.7+\y*1.4,3.5-\x *1.97) -- (5.6-\x*0.7+\y*1.4,2.4-\x *1.97) -- (6.4-\x*0.7+\y*1.4,2.4-\x *1.97) -- cycle;
   };
 \draw (0,-6) node[
                               text centered, 
                               color=white, 
                               font={\fontsize{8}{4}\sffamily\selectfont}
                             ] {FPAUO-#1/#2};
}} 
\makeatletter
\@addtoreset{equation}{section}
\makeatother

\begin{document}

~\vspace{-4cm}\begin{flushright}
\small
\FPAUO{10}{03}\\
IFT-UAM/CSIC-10-29\\
{\bf\tt arXiv:1006.0239}\\
June $1^{\rm st}$, $2010$
\normalsize
\end{flushright}
\begin{center}
\vspace{.5cm}
{\LARGE {\bf All the timelike supersymmetric}}\\[.5cm]
{\LARGE {\bf solutions of all ungauged $d=4$  supergravities}} 
\vspace{1.2cm}

 {\sl\large Patrick Meessen}
 \footnote{E-mail: {\tt meessenpatrick@uniovi.es}}$^{\dagger}$,
{\sl\large Tom{\'a}s Ort\'{\i}n}
\footnote{E-mail: {\tt Tomas.Ortin@cern.ch}}$^{\ddagger}$
 {\sl\large and Silvia Vaul\`a}
 \footnote{E-mail: {\tt Silvia.Vaula@uam.es}}$^{\ddagger}$

\vspace{.8cm}

$^{\dagger}${\it Department of Physics, University of Oviedo,\\
Avda. Calvo Sotelo s/n, E-33007 Oviedo, Spain}

\vspace{.3cm}

$^{\ddagger}${\it Instituto de F\'{\i}sica Te\'orica UAM/CSIC\\
Facultad de Ciencias C-XVI,  C.U.~Cantoblanco,  E-28049 Madrid, Spain}\\

\vspace{1.5cm}
{\large Dedicated to Prof.~Riccardo D'Auria on his 70th Birthday}
\vspace{1.5cm}


{\bf Abstract}

\end{center}

\begin{quotation}\small

  We determine the form of all timelike supersymmetric solutions of all $N\geq
  2, d=4$ ungauged supergravities, for $N\leq 4$ coupled to vector
  supermultiplets, using the $\mathrm{Usp}(\bar{n},\bar{n})$-symmetric
  formulation of Andrianopoli, D'Auria and Ferrara and the spinor-bilinears
  method, while preserving the global symmetries of the theories all the way.

  As previously conjectured in the literature, the supersymmetric solutions
  are always associated to a truncation to an $N=2$ theory that may include
  hypermultiplets, although fields which are eliminated in the truncations can
  have non-trivial values, as is required by the preservation of the global
  symmetry of the theories.

  The solutions are determined by a number of independent functions, harmonic
  in transverse space, which is twice the number of vector fields of the
  theory ($\bar{n}$). The transverse space is flat if an only if the would-be
  hyperscalars of the associated $N=2$ truncation are trivial.

\end{quotation}

\newpage

\pagestyle{plain}


\tableofcontents

\newpage

\section{Introduction}

The supersymmetric solutions of supergravity theories describing vacua, black
holes or topological defects, play a fundamental role in the progress of
superstring theory and related areas of research. It is, therefore, very
important to find and study as many supersymmetric solutions as possible, a
goal to which a huge effort has been devoted in the last few years. 

In his pioneering work \cite{Tod:1983pm}, Tod showed that it was possible to
systematically find all the supersymmetric configurations and solutions of a
given supergravity theory (pure $N=2,d=4$ in the case he considered, following
the lead of Ref.~\cite{Gibbons:1982fy}) by exploiting the consistency and
integrability conditions of the Killing spinor equations. He found that the
supersymmetric solutions of pure $N=2,d=4$ supergravity fall in two classes:
timelike and null. By \textit{all the supersymmetric configurations} we mean
all the field configurations that admit at least one Killing spinor, or equivalently one supercharge out of the $4N$ possible ones.
The timelike supersymmetric solutions are generalizations of the
Perj\`es-Israel-Wilson \cite{Perjes:1971gv} stationary solutions of the
Einstein-Maxwell system which themselves generalize the static solutions found by
Papapetrou and Majumdar \cite{Papaetrou:1947ib}. The solutions in the null
class are examples of Brinkmann waves \cite{Brinkmann:1925fr}.  Tod's feat
opened up the possibility of finding all the supersymmetric solutions of all
the supergravity theories.

Tod \cite{Tod:1983pm,Tod:1995jf} used the Newman-Penrose formalism to find the
supersymmetric solutions of the 4-dimensional pure $N=2$ and $4$ supergravity
theories, so that new techniques had to be developed in order to tackle
higher-dimensional cases.  In Ref.~\cite{Gauntlett:2002nw} Gauntlett {\it et
  al.\/} proposed to work with the spinor bilinears that can be constructed
out of the Killing spinors. These tensors satisfy a number of algebraic and
differential equations that follow from the Fierz identities and the original
Killing spinor equations that their constituents satisfy and which capture
enough (if not all the) information contained in them. The consistency and
integrability conditions of these new equations then determine the
supersymmetric configurations of the theory. In this way, in
Ref.~\cite{Gauntlett:2002nw} all the supersymmetric solutions of minimal
supergravity in $d=5$ dimensions were determined. These results were
immediately extended to the Abelian gauged case \cite{Gauntlett:2003fk} and
later on to general matter contents and couplings \cite{Bellorin:2006yr}
(always in the minimal $N=2$ supergravity). The spinor-bilinear method was
subsequently applied to other 4-dimensional
\cite{Caldarelli:2003pb}-\cite{Klemm:2010mc}, 6-dimensional
\cite{Gutowski:2003rg}, 7-dimensional \cite{Cariglia:2004qi}, 11-dimensional
\cite{Gauntlett:2002fz} and, recently, to 3-dimensional \cite{Deger:2010rb}
supergravities. 

In this approach (which will be used in this article) the form of all the field
configurations admitting at least one Killing spinor can be determined but
(unless further work is done) no classification of the supersymmetric
configurations by the number of independent Killing spinors they admit is
done. A different (but fundamentally equivalent) approach based on spinorial
geometry was developed in Refs.~\cite{Gillard:2004xq}. It has advantages over
the spinor-bilinear approach: using it, an exhaustive classification of the
configurations with different numbers of unbroken supersymmetries can be
achieved, also in higher dimensional theories where the application of 
the bilinear approach becomes unwieldy, by choosing convenient bases for the spinors.

Yet another approach, more adequate for finding supersymmetric solutions with
special geometries or properties, exploits the fact that a Killing spinor
defines a ``G structure''
\cite{Gauntlett:2002nw,Gauntlett:2002fz,Gauntlett:2002sc}.  Finally, another
approach used to find the timelike supersymmetric solutions of 4-dimensional
theories, and applied in particular to black holes, exploits the symmetries of
the dimensionally-reduced theories which become a non-linear $\sigma$-model
coupled to 3-dimensional gravity \cite{Bossard:2009at}. The main difficulty of
this powerful approach resides in the reconstruction of the 4-dimensional
solutions from the 3-dimensional ones.

The spinor-bilinear method that we are going to use is, we think, more adequate to find
large classes of solutions preserving (as a class) the global symmetries of
the theory: using it, it has been possible to find the general form of the
(pure, ungauged) $N=4,d=4$ supergravity black holes
\cite{Tod:1995jf,Bellorin:2005zc} written in an $\mathrm{SO}(6)$-covariant
form although some of them (which are singular), characterized by particular
choices of the charges, preserve $1/2$ of the supersymmetries instead of the
generic $1/4$ \cite{Bergshoeff:1996gg}. 

The spinor-bilinear method, however, becomes difficult to use for $N> 2$. For
instance, in the timelike $N=2$ case with one Killing spinor $\epsilon^{I}$
($I=1,2$) one can construct precisely four vector bilinears\footnote{See
  Appendix~\ref{sec-Fierz}.} $V^{I}{}_{J\, \mu} \equiv
i\bar{\epsilon}^{I}\gamma^{\mu}\epsilon_{J}$ which can be used as a tetrad to
construct the spacetime metric. For $N>2$ we have too many vector bilinears
and choosing four of them as a tetrad while preserving the $\mathrm{U}(N)$
invariance of the procedure seems impossible. There are several manifestations
of the same problem in the whole procedure.

Another problem, one that is common to all approaches, is the necessity of
treating different values of $N$ separately due to the different field content
and symmetries of each theory. 

In this paper we are going to use the spinor-bilinear method to determine the
general form of all the timelike supersymmetric solutions of all the $N\geq
2,d=4$ ungauged supergravities coupled to matter vector multiplets (when these
supermultiplets are available). As we will show, the main difficulties of the
spinor-method problem can be solved at least to the extent that the solution
allows us to determine the general form of all the timelike supersymmetric
solutions. This has required a deeper study of the algebra of spinor bilinears
than has been made in the literature hitherto and which has allowed us to find
a way to define an $\mathrm{SU}(2)$ subgroup without explicitly breaking the
$\mathrm{U}(N)$ R-symmetry of the equations.  Furthermore, we are going to use
the $N$-independent ``supergravity tensor calculus'' introduced in
Ref.~\cite{Andrianopoli:1996ve}, which allows the simultaneous study of all
the $N\geq 2,d=4$ ungauged supergravities just as one can work with tensors
constructed over vector spaces of undetermined number of dimensions and obtain
results valid for any $d$.

We have found that each timelike supersymmetric solutions is closely related
to a truncation to an $N=2$ theory determined by a $\mathrm{U}(2)$ subgroup of
the $\mathrm{U}(N)$ R-symmetry group\footnote{For supersymmetric black holes,
  this fact was conjectured in Ref.~\cite{Ferrara:2006yb} and earlier in
  Ref.~\cite{Andrianopoli:1997wi} and recently proven in the next to last of
  Refs.~\cite{Bossard:2009at}.}. It has to be emphasized that this
does not mean that each of them is just a solution of an $N=2$ truncation
since, for instance, all the vector fields are generically non-vanishing and
some of them would be eliminated by a generic truncation to $N=2$. However
most (if not all) of them may be generated by duality relations from a
solution of the associated $N=2$ truncation. This process can be rather
cumbersome but, in any case, our results render it unnecessary.

The construction of any timelike supersymmetric solution proceeds along the
following steps:

\begin{enumerate}

\item We have to choose the $\mathrm{U}(2)$ subgroup which determines the
  associated $N=2$ truncation:

\begin{enumerate}
\item Choose an $x$-dependent, rank-2, $N\times N$ complex antisymmetric
  matrix $M_{IJ}$ satisfying $M_{I[J}M_{KL]}=0$ ($x$ stands for the 3 spatial coordinates).  
  With it we can construct

\begin{displaymath}
  \mathcal{J}^{I}{}_{J} \equiv 2 |M|^{-2}M^{IK} M_{JK}\, ,
  \hspace{1.5cm}
  |M|^{2} = M^{PQ}M_{PQ}\, ,
\end{displaymath}

\noindent
which is a Hermitean projection operator whose trace is $+2$: $\mathcal{J}$
projects onto the above-mentioned $\mathrm{U}(2)$ subgroup. 

$\mathcal{J}$ must be covariantly constant\footnote{Naively one may think that
  it is always possible to choose a basis in $\mathrm{U}(N)$ space such that,
  for instance, $M_{12} = -M_{21} = +1$ and the rest of the components vanish,
  whence $\mathcal{J}$ is the identity in the corresponding 2-dimensional
  subspace. However, the necessary change of basis involves an, \textit{a
    priori}, arbitrary local $\mathrm{U}(N)$ rotation and the theory is not
  really $\mathrm{U}(N)$ gauge-invariant even if some fields undergo
  field-dependent compensating $\mathrm{U}(N)$ transformations when one
  performs a global symmetry transformation and there is a $\mathrm{U}(N)$
  gauge connection which is a composite field.

  This problem was first observed by Tod in his study of the $N=4$ theory
  \cite{Tod:1995jf} and, being unable to prove it, he conjectured that this
  rotation was always possible.

  We have not been able to prove this hypothesis in general either. We have
  proven that covariant constancy is required, though, which implies in the
  pure $N=4$ case studied by Tod ($\Omega^{I}{}_{J}\sim \delta^{I}{}_{J}$) as
  well as in the pure $N=3$ theory ($\Omega=0$) that $\mathcal{J}$ has to be
  constant.}

\begin{displaymath}
  \mathfrak{D}\mathcal{J}\equiv d\mathcal{J} -[\mathcal{J},\Omega ]=0\, ,  
\end{displaymath}

\noindent
in all cases.  In practice, the imposition of this requirement may be
postponed to the last stages of the construction of the supersymmetric
solutions.

Parametrizing the most general matrix $M_{IJ}$ that satisfies these
requirements gives a parametrization of the most general timelike
supersymmetric solutions.

\item Given $M_{IJ}$ and hence the covariantly-constant
  $\mathcal{J}^{I}{}_{J}$, we have to find three Hermitean, traceless,
  $x$-dependent $N\times N$ matrices $(\sigma^{m})^{I}{}_{J}$ ($m=1,2,3$),
  satisfying the same properties as the Pauli matrices in the subspace
  preserved by $\mathcal{J}$ as derived in App.~(\ref{sec-Fierz}), to wit

  \begin{displaymath}
    \begin{array}{rcl}
      \sigma^{m}\sigma^{n} 
      & = & 
      \delta^{mn}\mathcal{J} +i\varepsilon^{mnp}\sigma^{p}\, ,
      \\
      & &  \\
      \mathcal{J}\sigma^{m} 
      &  = &  
      \sigma^{m} \mathcal{J} = \sigma^{m}\, ,
      \\
      & &  \\
      \mathcal{J}^{K}{}_{J}\mathcal{J}^{L}{}_{I}
      & = & 
      \tfrac{1}{2}\mathcal{J}^{K}{}_{I}\mathcal{J}^{L}{}_{J}
      +
      \tfrac{1}{2}(\sigma^{m})^{K}{}_{I}(\sigma^{m})^{L}{}_{J}\, ,
      \\
      & &  \\
      M_{K[I}(\sigma^{m})^{K}{}_{J]} 
      & = & 0\, ,
      \\  
      & &  \\
      2|M|^{-2}M_{LI}(\sigma^{m})^{I}{}_{J}M^{JK} 
      & = & (\sigma^{m})^{K}{}_{L}\, .
      \\  
    \end{array}
  \end{displaymath}


  It turns out that we also have to impose the constraint

\begin{displaymath}
\mathcal{J}d\sigma^{m}\mathcal{J}=0\, , 
\end{displaymath}

\noindent
implying that the $\sigma$-matrices are constant in the subspace preserved by
the projector $\mathcal{J}$.\footnote{This is automatically satisfied for the
  projector itself $\mathcal{J}d\mathcal{J}\mathcal{J}=0$.}

\end{enumerate}

The four matrices $\{\mathcal{J},\sigma^{m}\}$ provide a
basis for the $\mathrm{U}(2)$ subgroup of the associated $N=2$ truncation and
can be seen as generators of its R-symmetry group.

Defining the complementary projector $\tilde{\mathcal{J}}\equiv
\mathbb{I}_{N\times N} -\mathcal{J}$ it is possible to separate the scalars
into those corresponding to the would-be vector multiplets and hypermultiplets
of the associated $N=2$ truncation. Thus, from the scalars in the generic
supergravity multiplet, described by the (pullback of the) Vielbein $P_{IJKL\,
  \mu}\equiv P_{[IJKL]\, \mu}$ and from the scalars in the generic matter
multiplet, described by $P_{i\, IJ\, \mu}\equiv P_{i\, [IJ]\, \mu}$; those in
the vector multiplets are described by

\begin{displaymath}
P_{IJKL}\,  \mathcal{J}^{I}{}_{[M}\mathcal{J}^{J}{}_{N} 
\tilde{\mathcal{J}}^{K}{}_{P}
  \tilde{\mathcal{J}}^{L}{}_{Q]}\,\,\,\,\,
\mbox{and}\,\,\,\,\,\,
P_{i\, IJ}\,   \mathcal{J}^{I}{}_{[K}\mathcal{J}^{J}{}_{L]}\, , 
\end{displaymath}

\noindent
and those in the hypermultiplets are described by

\begin{displaymath}
P_{IJKL}\, \mathcal{J}^{I}{}_{[M} \tilde{\mathcal{J}}^{J}{}_{N}
  \tilde{\mathcal{J}}^{K}{}_{P} \tilde{\mathcal{J}}^{L}{}_{Q]}
\,\,\,\,\,\,
\mbox{and}\,\,\,\,\,\,
P_{i\,  IJ}\, \mathcal{J}^{I}{}_{[K} \tilde{\mathcal{J}}^{J}{}_{L]}\, .
\end{displaymath}

The discrimination between these two kinds of scalars is, however, important:
those corresponding to the vector multiplets are sourced by the electric and
magnetic charges and enter into the attractor mechanism while those
corresponding to the hypermultiplets are not and should be frozen in
supersymmetric black-hole solutions.


\item Once the choice of $\mathrm{U}(2)$ subgroup is made, the solutions are
  constructed by the following procedure\footnote{This procedure is completely
    analogous to the procedure used to build supersymmetric solutions in
    ungauged $N=2$ theories coupled to vector multiplets and hypermultiplets
    described in Ref.~\cite{Huebscher:2006mr}}:

\begin{enumerate}

\item Using the symplectic functions of the scalars $\mathcal{V}_{IJ}$
  (\ref{eq:symsec}), which generalize the canonical symplectic section
  $\mathcal{V}$ of the $N=2$ theories \cite{deWit:1984pk}, we define the real
  symplectic vectors $\mathcal{R}$ and $\mathcal{I}$ by

  \begin{displaymath}
    \mathcal{R}+i\mathcal{I} \equiv  |M|^{-2}\mathcal{V}_{IJ}M^{IJ}\, ,
  \end{displaymath}

\noindent
which are $\mathrm{U}(N)$ singlets. No particular $\mathrm{U}(N)$ gauge-fixing
is necessary to construct the solutions.

\item For the supersymmetric solutions, the components of the symplectic
  vector $\mathcal{I}$ are real functions satisfying the Laplace equation in
  the 3-dimensional transverse space with metric
  $\gamma_{\underline{m}\underline{n}}$, to be described later. This is the
  only differential equation that needs to be solved.

\item $\mathcal{R}$ can in principle be found from $\mathcal{I}$ by solving
  the generalization of the so-called \textit{stabilization equations}.

\item The metric of the solutions has the form

\begin{displaymath}
  ds^{2} \; =\; |M|^{2} (dt+\omega)^{2} 
  -|M|^{-2}\gamma_{\underline{m}\underline{n}}dx^{m}dx^{n}\, .
\end{displaymath}

\noindent
where  

\begin{displaymath}
\begin{array}{rcl}
|M|^{-2} 
& = &  
\langle\,\mathcal{R}\mid \mathcal{I}\, \rangle\, ,
\\
& & \\
(d\omega)_{mn} 
& = & 
2\epsilon_{mnp}
\langle\,\mathcal{I}\mid \partial^{p}\mathcal{I}\, \rangle\, ,
\end{array}
\end{displaymath}

\noindent
so they can be computed directly from $\mathcal{R}$ and $\mathcal{I}$. 

The 3-dimensional transverse metric $\gamma_{\underline{m}\underline{n}}$ is
determined indirectly by the would-be hypers; in particular, when those
scalars are frozen the metric is flat.  The full condition that the
3-dimensional metric has to satisfy is that its spin-connection must be
related to (the pullback of) the connection of the scalar manifold, $\Omega$ in
(\ref{eq:O}), by

\begin{displaymath}
\varpi^{mn} = i\varepsilon^{mnp}\mathrm{Tr}\, [\sigma^{p}\Omega]\, .
\end{displaymath}

\noindent
Observe that only the $\mathfrak{su}(2)$ part of $\Omega$ contributes to
$\varpi^{mn}$\footnote{It plays the same r\^{o}le as the $\mathfrak{su}(2)$
  connection of the hyper-K\"ahler manifold in Ref. $\cite{Huebscher:2006mr}$
  and the condition on the metric is identical to the one found in the $N=2$
  case although in that case the $2\times 2$ matrices $\sigma^{m}$ are the
  standard, constant, Pauli matrices}.

\item The vector field strengths are given by 

\begin{displaymath}
F=  -{\textstyle\frac{1}{2}} d (\mathcal{R}\hat{V})   
-{\textstyle\frac{1}{2}}\star
(\hat{V}\wedge 
d\mathcal{I}) 
\, ,  
\hspace{1.5cm}
\hat{V} = \sqrt{2}|M|^{2}(dt+\omega)\, . 
\end{displaymath}

\item The scalars corresponding to the vector multiplets in the associated
  $N=2$ truncation, represented by the projected Vielbeine

\begin{displaymath}
P_{IJKL}\,  \mathcal{J}^{I}{}_{[M}\mathcal{J}^{J}{}_{N} \tilde{\mathcal{J}}^{K}{}_{P}
  \tilde{\mathcal{J}}^{L}{}_{Q]}\,\,\,\,\,\,
\mbox{and}\,\,\,\,\,\,
P_{i\, IJ}\,   \mathcal{J}^{I}{}_{[K}\mathcal{J}^{J}{}_{L]}\, , 
\end{displaymath}

\noindent
can in principle be found from $\mathcal{R}$ and $\mathcal{I}$. The Killing
Spinor Identities guarantee that the equations of motion of these scalars are
satisfied if the Maxwell equations and Bianchi identities are
satisfied\footnote{Actually, the only independent equations of motion that
  need to be solved are the $0^{th}$ components of the Maxwell equations and
  Bianchi identities. Some of the other equations are just automatically
  satisfied for supersymmetric configurations and the rest is proportional to
  those $0^{th}$ components.}, which is the case when the components of
$\mathcal{I}$ are harmonic functions on the transverse space.

\item The scalars corresponding to the hypers, described by the Vielbeine

\begin{displaymath}
P_{IJKL}\, \mathcal{J}^{I}{}_{[M} \tilde{\mathcal{J}}^{J}{}_{N}
  \tilde{\mathcal{J}}^{K}{}_{P} \tilde{\mathcal{J}}^{L}{}_{Q]}
\,\,\,\,\,\,
\mbox{and}\,\,\,\,\,\,
P_{i\,  IJ}\, \mathcal{J}^{I}{}_{[K} \tilde{\mathcal{J}}^{J}{}_{L]}\, ,
\end{displaymath}

\noindent
must be found independently by solving the supersymmetry constraints

\begin{displaymath}
\begin{array}{rcl}
P_{IJKL\, m}\, \mathcal{J}^{I}{}_{[M}
\tilde{\mathcal{J}}^{J}{}_{N} \tilde{\mathcal{J}}^{K}{}_{P}
\tilde{\mathcal{J}}^{L}{}_{Q]} (\sigma^{m})^{Q}{}_{R} 
& = & 
0\, ,
\\
& & \\  
P_{i\, IJ\, m}\, \mathcal{J}^{I}{}_{[K}
\tilde{\mathcal{J}}^{J}{}_{L]}(\sigma^{m})^{L}{}_{M} 
& = & 
0\, .
\\
\end{array}
\end{displaymath}

\noindent
The Killing Spinor Identities guarantee that their equations of motion are
automatically solved\footnote{This situation is completely analogous to what
  happens with the hyperscalars of $N=2$ theories \cite{Huebscher:2006mr}}.

\end{enumerate}

\end{enumerate}

In the rest of this paper we are going to prove in full detail the above
result.  We are going to start by giving the generic description of all the
$N\geq 2, d=4$ supergravities with vector multiplets (where available) in
Section~\ref{sec-ngeq2sugras}. In Section~\ref{sec-kses} we are going to
present the Killing spinor equations for all these theories and we are going
to find the Killing Spinor Identities that constrain the off-shell equations
of motion of the bosonic fields for supersymmetric field configurations.


\section{Generic description of $N\geq 2, d=4$ Supergravities}
\label{sec-ngeq2sugras}

We are going to study all the $N\geq 2, d=4$ supergravities coupled to vector
multiplets simultaneously, using the fact that all the supergravity multiplets
and all the vector multiplets for all $N=1, \cdots, 8$ can be written in the
same generic form \cite{Andrianopoli:1996ve}; we only need to take into
account the range of values taken by the $\mathrm{U}(N)$ R-symmetry indices,
denoted by uppercase Latin letters $I$ \textit{etc.} taking on values $1,
\cdots ,N$, in each particular case\footnote{This formalism is taken from
  Ref.~\cite{Andrianopoli:1996ve}, but adapted to the notations of
  Ref.~\cite{Bellorin:2005zc}.  Furthermore, throughout this paper we use the
  convention that the only fields and terms that should be considered are
  those whose number of antisymmetric $\mathrm{SU}(N)$ indices is correct,
  i.e.~objects with more than $N$ antisymmetric indices are zero and terms
  with Levi-Civit\`a symbols $\epsilon^{I_{1}\cdots I_{M}}$ should only be
  considered when $M$ equals the $N$ of the supergravity theory under
  consideration. There are also constraints on the generic fields for specific
  values of $N$ that we are going to review.}.

The generic supergravity multiplet in four dimensions is

\begin{equation}
\label{Ngrav}
\left\{ e^{a}{}_{\mu},\psi_{I\, \mu},A^{IJ}{}_{\mu},\chi_{IJK},
\chi^{IJKLM},\,P_{IJKL\, \mu}\right\}\, ,\,\,\,\,
I,J,\dots=1,\cdots, N\, ,
\end{equation}

\noindent
and the generic vector multiplets (labeled by $i=1,\cdots,n$) are

\begin{equation}
\left\{A_{i\, \mu},\lambda_{iI},\lambda_{i}{}^{IJK},P_{iIJ\, \mu}\right\}\, .
\end{equation}

\noindent
The spinor fields $\psi_{I\, \mu},\chi_{IJK},\chi^{IJKLM},
\lambda_{iI},\lambda_{i}{}^{IJK}$ have positive chirality with the given
positions of the $\mathrm{SU}(N)$ indices. 

The scalars of these theories are encoded into the $2\bar{n}$-dimensional
($\bar{n}\equiv n+\frac{N(N-1)}{2}$) symplectic vectors ($\Lambda=1,\dots
\bar{n}$) $\mathcal{V}_{IJ}$ and $\mathcal{V}_{i}$ whose properties are
reviewed in Appendix~\ref{sec-scalars}. They appear in the bosonic sector of
the theory via the pullbacks of the Vielbeine $P_{IJKL\mu}$ (supergravity
multiplet) and $P_{iIJ\, \mu}$ (matter multiplets)\footnote{The Vielbeine
  $P_{ij\, \mu}$ either vanish identically or depend on $P_{IJKL\mu}$ and
  $P_{iIJ\, \mu}$, depending on the specific value of $N$. Thus, they are not
  needed as independent variables to construct the theories.}.  There are
three instances of theories for which the scalar Vielbeine are constrained:
first, when $N=4$ the matter scalar Vielbeine are constrained by the
$\mathrm{SU}(4)$ complex self-duality relation\footnote{ In order to highlight
  the fact that an equation holds for a specific $N$ only, we write a
  numerical variation of the token ``$N=4::$'' to the left of the
  equation.  }

\begin{equation}
N\ =\ 4 ::\hspace{.6cm}
P^{*\, i\, IJ} \ =\  \tfrac{1}{2}\varepsilon^{IJKL}\ P_{i\, KL}\, .
\end{equation}

Secondly, in $N=6$ the scalars in the supergravity multiplet are represented
by one Vielbein $P_{IJ}$ and one Vielbein $P_{IJKL}$ related by the
$\mathrm{SU}(6)$ duality relation

\begin{equation}
N\ =\ 6 ::\hspace{.6cm}
P^{*\, IJ} \ =\ \tfrac{1}{4!}\varepsilon^{IJK_{1}\cdots K_{4}} \ P_{K_{1}\cdots K_{4}}\, ,
\end{equation}

\noindent
and lastly the $N=8$ case, in which the Vielbeine is constrained by the
$\mathrm{SU}(8)$ complex self-duality relation

\begin{equation}
N\ =\ 8 ::\hspace{.6cm}
P^{*\, I_{1}\cdots I_{4}} =\tfrac{1}{4!}\varepsilon^{I_{1}\cdots
  I_{4}J_{1}\cdots J_{4}} \ P_{J_{1}\cdots J_{4}}\, .
\end{equation}

\noindent
These constraints must be taken into account in the action.

The graviphotons $A^{IJ}{}_{\mu}$ do not appear directly in the theory, rather
they only appear through the ``dressed'' vectors, which are defined by

\begin{equation} 
A^{\Lambda}{}_{\mu}\equiv 
{\textstyle\frac{1}{2}}f^{\Lambda}{}_{IJ}A^{IJ}{}_{\mu}
+f^{\Lambda}{}_{i} A^{i}{}_{\mu}\, .
\end{equation}

The action for the bosonic fields is

\begin{equation}
  \begin{array}{rcl}
S & = & {\displaystyle\int} d^{4}x\sqrt{|g|}
\left[
R
+2\Im{\rm m}\mathcal{N}_{\Lambda\Sigma} 
F^{\Lambda\, \mu\nu}F^{\Sigma}{}_{\mu\nu}
-2\Re{\rm e}\mathcal{N}_{\Lambda\Sigma} 
F^{\Lambda\, \mu\nu}\star F^{\Sigma}{}_{\mu\nu}
\right.
\\
& & \\
& &
\hspace{2cm}
\left.
+\frac{2}{4!}\alpha_{1}P^{*\, IJKL}{}_{\mu}P_{IJKL}{}^{\mu}
+\alpha_{2}P^{*\, iIJ}{}_{\mu}P_{iIJ}{}^{\mu}
\right]\, ,
\end{array}
\end{equation}

\noindent
where $\mathcal{N}_{\Lambda\Sigma}$ is the generalization of the $N=2$ period
matrix, defined in Eq.~(\ref{eq:periodmatrix}), and where the parameters
$\alpha_{1},\alpha_{2}$ are equal to $1$ in all cases except for $N=4,6$ and
$8$ as one needs to take into account the above constraints on the Vielbeine:
$\alpha_{2}=1/2$ for $N=4$, $\alpha_{1}+\alpha_{2}=1$ for $N=6$ (the simplest
choice being $\alpha_{2}=0$) and $\alpha_{1}=1/2$ for $N=8$. The action is
good enough to compute the Einstein and Maxwell equations, but not the
scalars' equations of motion in the cases in which the scalar Vielbeine are
constrained: these constraints have to be properly dealt with and the
resulting equations of motion are given below.
 
The supersymmetry transformations of the bosonic fields can be written in the
form

\begin{eqnarray}
\delta_{\epsilon} e^{a}{}_{\mu} 
& = & 
-i\bar{\psi}_{I\mu}\gamma^{a}\epsilon^{I}
-i\bar{\psi}^{I}{}_{\mu}\gamma^{a}\epsilon_{I}\, ,
\\
& & \nonumber \\
\delta_{\epsilon} A^{\Lambda}{}_{\mu} 
& = & 
f^{\Lambda}{}_{IJ}\bar{\psi}^{I}{}_{\mu}\epsilon^{J}
+f^{*\Lambda  IJ}\bar{\psi}_{I\mu}\epsilon_{J}
-{\textstyle\frac{i}{2}}
(f^{\Lambda}{}_{i}\bar{\lambda}^{iI}\gamma_{\mu}\epsilon_{I}
+f^{*\, \Lambda i}\bar{\lambda}_{iI}\gamma_{\mu}\epsilon^{I})
\nonumber \\
& & \nonumber \\
& &
-{\textstyle\frac{i}{4}}
(f^{\Lambda}{}_{IJ}\bar{\chi}^{IJK}\gamma_{\mu}\epsilon_{K}
+f^{*\Lambda  IJ}\bar{\chi}_{IJK}\gamma_{\mu}\epsilon^{K})\, ,
\\
& & \nonumber \\
(U^{-1}\delta_{\epsilon} U)_{IJKL}  
& = &
4\bar{\chi}_{[IJK}\epsilon_{L]} +\bar{\chi}_{IJKLM}\epsilon^{M}\, ,
\\
& & \nonumber \\
(U^{-1}\delta_{\epsilon} U)_{iIJ} 
& = &
2\bar{\lambda}_{i[I}\epsilon_{J]}
+{\textstyle\frac{1}{2}}\bar{\lambda}_{iIJK}\epsilon^{K}\, ,
\end{eqnarray}

\noindent
where $U$ is the $\mathrm{Usp}(\bar{n},\bar{n})$ matrix describing the
scalars, defined in Eq.~(\ref{eq:U}). Those of the fermionic fields can be put
in the form

\begin{eqnarray}
\label{eq:dp}
\delta_{\epsilon} \psi_{I\mu} 
& = &
\mathfrak{D}_{\mu}\epsilon_{I}+T_{IJ}{}^{+}{}_{\mu\nu}\gamma^{\nu}
\epsilon^{J}\, ,
\\ 
& & \nonumber \\
\label{eq:dxijk}
\delta_{\epsilon} \chi_{IJK} 
& = & 
-\tfrac{3i}{2} \not\!T_{[IJ}{}^{+}\epsilon_{K]}
+i\not\!P_{IJKL}\epsilon^{L}\, ,
\\
& & \nonumber \\
\label{eq:dlii}
\delta_{\epsilon}\lambda_{iI}  
& = &  
-\tfrac{i}{2} \not\!T_{i}{}^{+}\epsilon_{I}
+i\not\!P_{iIJ}\epsilon^{J}\, ,
\\
& & \nonumber \\
\label{eq:dxijklm}
\delta_{\epsilon} \chi_{IJKLM} 
& = &
-5i \not\!P_{[IJKL}\epsilon_{M]}
+\tfrac{i}{2}\varepsilon_{IJKLMN} \not\!T^{-}\epsilon^{N}
+\tfrac{i}{4}\varepsilon_{IJKLMNOP} \not\!T^{NO-}\epsilon^{P}\, ,\\
& & \nonumber \\
\label{eq:dliijk}
\delta_{\epsilon} \lambda_{iIJK} & = &
-3i \not\!P_{i[IJ}\epsilon_{K]}
+\tfrac{i}{2} \varepsilon_{IJKL} \not\!T_{i}{}^{-}\epsilon^{L}
+\tfrac{i}{4}\varepsilon_{IJKLMN}\not\!T^{LM-}\epsilon_{N}\, ,
\end{eqnarray}

\noindent 
where we have defined the graviphoton and matter vector field strengths

\begin{equation} 
T_{IJ}{}^{+}{}_{\mu\nu}= 2if^{\Lambda}{}_{IJ}
\ \Im{\rm m}\mathcal{N}_{\Lambda\Sigma}\ F^{\Sigma+}{}_{\mu\nu}\, ,\,\,\,\,
T_{i}{}^{+}{}_{\mu\nu}=2if^{\Lambda}{}_{i} 
\ \Im{\rm m}\mathcal{N}_{\Lambda\Sigma}\ F^{\Sigma+}{}_{\mu\nu}\, ,
\end{equation}

\noindent
and where 

\begin{equation}
\mathfrak{D}_{\mu}\epsilon_{I} \equiv \nabla_{\mu}\epsilon_{I}
-\epsilon_{J}\Omega_{\mu}{}^{J}{}_{I}\, ,  
\end{equation}

\noindent
$\Omega$ being the pullback of the connection on the scalar manifold, defined
in Appendix~\ref{sec-scalars}.

We stress that, according to our conventions, the terms with
$\varepsilon$-symbols should only be considered when the value of $N$ equals
its rank.  {}Furthermore, when $N=4,6$ or $8$ Eqs.~(\ref{eq:dxijklm}) and
(\ref{eq:dliijk}) depend on the first three supersymmetry rules, whereas for
$N=2$ they are equations for non-existing fields: therefore,
Eqs.~(\ref{eq:dxijklm}) and (\ref{eq:dliijk}) only need to be considered in
the cases $N=3$ and $5$, and then only the first term on the l.h.s.~is
non-vanishing.

For convenience, we denote the Bianchi identities for the vector field
strengths by

\begin{equation}
\label{eq:BL}
\mathcal{B}^{\Lambda\, \mu} \equiv \nabla_{\nu}\star F^{\Lambda\,
  \nu\mu}\, .  
\end{equation}

\noindent
and the bosonic equations of motion by

\begin{equation}
  \begin{array}{rclrcl}
\mathcal{E}_{a}{}^{\mu} &  \equiv &
-\frac{1}{2\sqrt{|g|}}{\displaystyle\frac{\delta S}{\delta e^{a}{}_{\mu}}}\, ,
\hspace{.5cm} &
\mathcal{E}^{IJKL} & \equiv &  -\frac{1}{2\sqrt{|g|}}
{\displaystyle\left(\frac{\delta S}{\delta U}U\right)^{IJKL}}
=-\frac{1}{2\sqrt{|g|}}P^{*\, IJKL\, A}
{\displaystyle\frac{\delta S}{\delta \phi^{A}}}\, ,
\\
& & & & & \\
\mathcal{E}_{\Lambda}{}^{\mu} & \equiv & \frac{1}{8\sqrt{|g|}}
{\displaystyle\frac{\delta S}{\delta A^{\Lambda}{}_{\mu}}}\, ,
\hspace{.5cm} & 
\mathcal{E}^{iIJ} & \equiv & -\frac{1}{2\sqrt{|g|}}
{\displaystyle\left(\frac{\delta S}{\delta U}U\right)^{iIJ}}
= -\frac{1}{2\sqrt{|g|}}P^{*\, iIJ\, A}
{\displaystyle\frac{\delta S}{\delta \phi^{A}}}\, ,\\
\end{array}
\end{equation}

\noindent
where $P^{*\, IJKL\, A}$ and $P^{*\, iIJ\, A}$ are the inverse Vielbeine and
$\phi^{A}$ are the physical fields of the theory.

The explicit forms of the Einstein and Maxwell equations are

\begin{eqnarray}
\mathcal{E}_{\mu\nu} & = & 
G_{\mu\nu}
+{\textstyle\frac{1}{12}}\alpha_{1}\left[P^{*\, IJKL}{}_{(\mu|}P_{IJKL\ | \nu)}
-{\textstyle\frac{1}{2}}g_{\mu\nu}
P^{*\, IJKL}{}_{\rho}P_{IJKL}{}^{\rho}\right]\nonumber \\
& & \nonumber \\
& & 
+\alpha_{2}P^{*\, iIJ}{}_{(\mu|}P_{iIJ\ | \nu)}
-{\textstyle\frac{1}{2}}g_{\mu\nu}
P^{*\, iIJ}{}_{\rho}P_{iIJ}{}^{\rho}
+8\Im {\rm m}\mathcal{N}_{\Lambda\Sigma}
F^{\Lambda\, +}{}_{\mu}{}^{\rho}F^{\Sigma\, -}{}_{\nu\rho}\, ,
\label{eq:Emn}\\
& & \nonumber \\
\mathcal{E}_{\Lambda}{}^{\mu} & = & 
\nabla_{\nu}\star\tilde{F}_{\Lambda}{}^{\nu\mu}\, ,
\label{eq:ERm} 
\end{eqnarray}

\noindent
where we have defined the dual vector field strength $\tilde{F}_{\Lambda}$ by

\begin{equation}
\tilde{F}_{\Lambda\, \mu\nu} \equiv   
-{\textstyle\frac{1}{4\sqrt{|g|}}}
\frac{\delta S}{\delta {}^{\star}F^{\Lambda}{}_{\mu\nu}}
= 2  \Re {\rm e}(\mathcal{N}_{\Lambda\Sigma} F^{\Sigma\, +})= 
\Re {\rm e}\mathcal{N}_{\Lambda\Sigma}F^{\Sigma}{}_{\mu\nu}
+\Im {\rm m}\mathcal{N}_{\Lambda\Sigma}\star F^{\Sigma}{}_{\mu\nu}\, .
\end{equation}

Using Eqs.~(\ref{eq:PIJKLdN}) and (\ref{eq:PiIJdN}) and taking into account
the constraints satisfied by the Vielbeine in the cases $N=4,6$ and $8$, we find
that the scalar equations of motion take the following forms, slightly
different for each value of $N$:

\begin{description}
\item[$N=2$::]

  \begin{equation}
    \label{eq:EiIJN=2}  
    \mathcal{E}^{iIJ}  =  \mathfrak{D}^{\mu}P^{*\, iIJ}{}_{\mu}
    + 2T^{i\, -}{}_{\mu\nu}T^{IJ\, -\, \mu\nu}
    +P^{*\, iIJ\, A}P^{*\, jk}{}_{A}T_{j}{}^{+}{}_{\mu\nu}
    T_{k}{}^{+\, \mu\nu}\, .
  \end{equation}

\item[$N=3$::]

\begin{equation}
  \label{eq:EiIJN=3}  
  \mathcal{E}^{iIJ}  =  \mathfrak{D}^{\mu}P^{*\, iIJ}{}_{\mu}
  + 2T^{i\, -}{}_{\mu\nu}T^{IJ\, -\, \mu\nu}\, .
\end{equation}

\item[$N=4$::]

\begin{eqnarray}
  \label{eq:EIJKLN=4}
  \mathcal{E}^{IJKL} & = & \mathfrak{D}^{\mu}P^{*\, IJKL}{}_{\mu}
  +6  T^{[IJ|-}{}_{\mu\nu}T^{|KL]-\, \mu\nu}
  +P^{*\, IJKL\, A}P^{*\, ij}{}_{A}T_{i}{}^{+}{}_{\mu\nu}
  T_{j}{}^{+\, \mu\nu}\, ,
  \\
  & & \nonumber \\
  \label{eq:EiIJN=4}
  \mathcal{E}^{iIJ} & = & \mathfrak{D}^{\mu}P^{*\, iIJ}{}_{\mu}
  + T^{i\, -}{}_{\mu\nu}T^{IJ\, -\, \mu\nu}
  +\tfrac{1}{2}\varepsilon^{IJKL}
  T_{i}{}^{+}{}_{\mu\nu}T_{KL}{}^{+\, \mu\nu}\, .
\end{eqnarray}

\item[$N=5$::] 

\begin{equation}
  \label{eq:EIJKLN=5}
  \mathcal{E}^{IJKL} = \mathfrak{D}^{\mu}P^{*\, IJKL}{}_{\mu}
  +6  T^{[IJ|-}{}_{\mu\nu}T^{|KL]-\, \mu\nu}\, .
\end{equation}

\item[$N=6$::]

\begin{equation}
  \label{eq:EIJKLN=6}
  \mathcal{E}^{IJKL}  = \mathfrak{D}^{\mu}P^{*\, IJKL}{}_{\mu}
  +6  T^{[IJ|-}{}_{\mu\nu}T^{|KL]-\, \mu\nu}
  +\varepsilon^{IJKLMN}T^{+}{}_{\mu\nu}T_{MN}{}^{+\, \mu\nu}
  \, .
\end{equation}

\item[$N=8$::]

\begin{equation}
  \label{eq:EIJKLN=8}
  \mathcal{E}^{IJKL}  = \mathfrak{D}^{\mu}P^{*\, IJKL}{}_{\mu}
  +6 T^{[IJ|-}{}_{\mu\nu}T^{|KL]-\, \mu\nu}
    +\tfrac{1}{4}\varepsilon^{IJKLMNPQ}T_{MN}{}^{+}{}_{\mu\nu}T_{PQ}{}^{+\,
      \mu\nu}\, .
\end{equation}

\end{description}

\section{Generic $N\geq 2, d=4$ Killing Spinor Equations and Identities}
\label{sec-kses}

The Killing spinor equations are

\begin{eqnarray}
\label{eq:kse1}
\mathfrak{D}_{\mu}\epsilon_{I}+T_{IJ}{}^{+}{}_{\mu\nu}\gamma^{\nu}
\epsilon^{J} & = & 0\, ,\\ 
& & \nonumber \\
\label{eq:kse2}
\not\!P_{IJKL}\epsilon^{L}
-{\textstyle\frac{3}{2}} \not \!T_{[IJ}{}^{+}\epsilon_{K]}
& = & 0\, ,\\
& & \nonumber \\
\label{eq:kse4}
\not\!P_{i\, IJ}\epsilon^{J}
-{\textstyle\frac{1}{2}} \not \!T_{i}{}^{+}\epsilon_{I}
& = & 0\, ,\\
& & \nonumber \\
\label{eq:kse3}
N=5\hspace{.2cm}::\hspace{2.5cm}
\not\!P_{[IJKL}\epsilon_{M]}
& = & 0\, ,\\
& & \nonumber \\
\label{eq:kse5}
N=3\hspace{.2cm}::\hspace{2.9cm}
\not\!P_{i\, [IJ}\epsilon_{K]}
& = & 0\, ,
\end{eqnarray}

\noindent
where, as indicated by the notation, the last two KSEs should only be
considered for $N=5$ and $N=3$, respectively.

From the bosonic supersymmetry transformation rules we immediately find using
the algorithm of Refs.~\cite{Kallosh:1993wx,Bellorin:2005hy}

\begin{eqnarray}
\label{eq:ksi1}
\mathcal{E}_{a}{}^{\mu}\gamma^{a}\epsilon^{I} 
-4i\mathcal{E}_{\Lambda}{}^{\mu}f^{*\, \Lambda\, IJ}\epsilon_{J}
& = & 0\, ,\\
& & \nonumber \\   
\label{eq:ksi2}
\mathcal{E}_{\Lambda}{}^{\mu}f^{*\, \Lambda\, [IJ}\gamma_{\mu}\epsilon^{K]}
-{\textstyle\frac{i}{3!}} 
\mathcal{E}^{IJKL}\epsilon_{L} & = & 0\, ,\\
& & \nonumber \\   
\label{eq:ksi3}
\mathcal{E}_{\Lambda}{}^{\mu}f^{*\, \Lambda\, i}\gamma_{\mu}\epsilon^{I}
-{\textstyle\frac{i}{2}} 
\mathcal{E}^{i\, IJ}\epsilon_{J} & = & 0\, ,\\
& & \nonumber \\   
\label{eq:ksi4}
N=5\hspace{.2cm}::\hspace{2.5cm}
\mathcal{E}^{[IJKL}\epsilon^{M]} & = & 0\, ,\\
& & \nonumber \\   
\label{eq:ksi5}
N=3\hspace{.2cm}::\hspace{2.9cm}
\mathcal{E}^{i\, [IJ}\epsilon^{K]} & = & 0\, .
\end{eqnarray}


In these equations it is implicitly assumed that the Bianchi identities are
satisfied, \textit{i.e.} $\mathcal{B}^{\Lambda\, \mu}=0$. It is, however,
convenient not to make use of this assumption as to preserve the manifest
electric-magnetic duality of the formalism. We can, and will, introduce the
Bianchi identities into these equations by the replacement

\begin{equation}
\mathcal{E}_{\Lambda}{}^{\mu}f^{\Lambda}\,\,\, \longrightarrow\,\,\,
\langle\,\mathcal{E} \mid \mathcal{V}\, \rangle\, , 
\end{equation}

\noindent
where $\mathcal{E}$ is the symplectic vector containing the Maxwell
equations and Bianchi identities.

We can start to derive consequences from these identities in terms of the
spinor bilinears defined and studied in Appendix~\ref{sec-Fierz} and in this
paper we will only study the case in which the vector bilinear, $V^{a}= i
\bar{\epsilon}^{I}\gamma^{a}\epsilon_{I}$, is timelike
($V^{2}=V^{a}V_{a}=2|M|^{2}>0$).


\subsection{Timelike case}
\label{sec-timelikeksis}

It is convenient to work with flat indices and use a Vierbein basis in which
$e^{0}\equiv \tfrac{1}{\sqrt{2}}|M|^{-1}V_{\mu}dx^{\mu}$.  Acting with
$i\bar{\epsilon}_{I}$ and $\bar{\epsilon}^{K}\gamma^{\nu}$ on the first KSI
Eq.~(\ref{eq:ksi1}) we get, 

\begin{eqnarray}
\label{eq:ksi1-1}
V^{b}\mathcal{E}_{b}{}^{a}+4
\langle\,\mathcal{E}^{a} \mid \mathcal{V}^{*\, IJ}\, \rangle M_{IJ} 
& = & 0\, , \\  
& & \nonumber \\   
\label{eq:ksi1-2}
\mathcal{E}_{c}{}^{a}(g^{cb}M^{KI} +\Phi^{KI\, cb}) +4
\langle\,\mathcal{E}^{a} \mid \mathcal{V}^{*\, JI}\, \rangle  
V^{K}{}_{J}{}^{b} & = & 0\, ,
\end{eqnarray}

\noindent
respectively. 
Multiplying the second identity with $M_{KI}$ we obtain

\begin{equation}
|M|^{2}\mathcal{E}^{ab} +2 \langle\,\mathcal{E}^{a} \mid
 \mathcal{V}^{*\, IJ}\, \rangle M_{IJ} V^{b} =0\, . 
\end{equation}

\noindent 
The symmetry and reality of the Einstein equation imply, firstly 

\begin{equation}
\label{eq:timelikeksi1}
\mathcal{E}^{0m} = \mathcal{E}^{mn} =0\, ,
\end{equation}

\noindent
so all components of the Einstein equations but $\mathcal{E}^{00}$ are
automatically and identically satisfied\footnote{As explained in
  Ref.~\cite{Bellorin:2006xr} this poses strong constraints on the sources of
  the solutions because having supersymmetry unbroken everywhere implies that
  the KSIs should be identically (i.e.~not up to $\delta$-function terms)
  satisfied everywhere.}; secondly\footnote{The imaginary part of the equation
  $\langle\,\mathcal{E}^{0} \mid \mathcal{I}\, \rangle =0$ is related to the
  absence of sources of NUT charge in globally supersymmetric solutions
  \cite{Bellorin:2006xr}.}

\begin{equation}
\label{eq:timelikeksi2-1}
\mathcal{E}^{00} = 
-2\sqrt{2}|M| \langle\,\mathcal{E}^{0} \mid \mathcal{R}\, \rangle \, ,
\end{equation}

\noindent
where we have defined the $\mathrm{U}(N)$-neutral real symplectic vectors $\mathcal{R}$
and $\mathcal{I}$ by

\begin{equation}
\label{eq:RandIdef}
|M|^{-2}M^{IJ} \mathcal{V}_{IJ} 
\equiv\mathcal{V}= \mathcal{R}+i\mathcal{I}\, ,
\end{equation}

\noindent
whence the remaining component of the Einstein equations is satisfied if the
$0^{th}$ component of the Maxwell equations and Bianchi identities are
satisfied.  Thirdly and finally

\begin{eqnarray}
\langle\,\mathcal{E}^{m} \mid \mathcal{R}\, \rangle  
& = & 0\, , \\
& & \nonumber \\
\label{eq:timelikeksi2-2}
\langle\,\mathcal{E}^{a} \mid \mathcal{I}\, \rangle
& = & 0\, .
\end{eqnarray}

Acting with $i\bar{\epsilon}_{L}$ and $\bar{\epsilon}^{L}\gamma^{\nu}$ on
Eq.~(\ref{eq:ksi2}), which is only to be considered for $N\geq 3$, we obtain

\begin{eqnarray}
\label{eq:ksi2-1}
\langle\,\mathcal{E}^{a} \mid \mathcal{V}^{*\, [IJ}\, \rangle  
V^{K]}{}_{L\ a}  
-{\textstyle\frac{1}{3!}} \mathcal{E}^{IJKM}M_{ML}
& = & 0\, ,\\
& & \nonumber \\
\label{eq:ksi2-2}
\langle\,\mathcal{E}^{a} \mid \mathcal{V}^{*\, [IJ}\, \rangle  
(-\delta^{b}{}_{a}M^{K]L} +\Phi^{K]L\ b}{}_{a} )
-{\textstyle\frac{1}{3!}} \mathcal{E}^{IJKM}V^{L\, b}{}_{M} 
& = & 0\, .
\end{eqnarray}

Multiplying Eq.~(\ref{eq:ksi2-1}) by $2M^{NL}|M|^{-2}$ and antisymmetrizing
the  four free indices we get 

\begin{equation}
\label{eq:timelikeksi4}
\langle\,\mathcal{E}^{a} \mid \mathcal{V}^{*\, [IJ}\, \rangle  
\frac{M^{KL]}}{|M|}
-{\textstyle\frac{1}{\sqrt{2}\cdot 3!}}\delta^{a}{}_{0}
 \mathcal{E}^{M[IJK}\mathcal{J}^{L]}{}_{M} 
= 0\, .
\end{equation}

\noindent
Setting $K=L$ in Eq.~(\ref{eq:ksi2-1}),
using the antisymmetric part of Eq.~(\ref{eq:ksi1-2}) and taking into account
Eq.~(\ref{eq:timelikeksi2-1}), we get

\begin{equation}
\label{eq:otraksi}
\langle\,\mathcal{E}^{m} \mid \mathcal{V}^{*\, IJ}\, \rangle 
=  
0\, ,
\end{equation}

\noindent
and 

\begin{equation}
\mathcal{E}^{IJKM}M_{KM}
=
-2\sqrt{2} |M|(\delta^{IJ}{}_{KL} -|M|^{-2}M^{IJ}M_{KL})
\langle\,\mathcal{E}^{0} \mid \mathcal{V}^{*\, KL}\, \rangle  \, .
\end{equation}

\noindent
This implies that the projections 

\begin{equation}
\mathcal{E}^{MNPQ}\, \mathcal{J}^{[I}{}_{M} \mathcal{J}^{J}{}_{N}
\tilde{\mathcal{J}}^{K}{}_{P} \tilde{\mathcal{J}}^{L]}{}_{Q}\, ,  
\end{equation}

\noindent
which should be understood as the equations of motion of the scalars that
would correspond to the vector multiplets scalars in the associated $N=2$
truncations, are satisfied if the $0^{th}$ component of the Maxwell equations
and Bianchi identities are.  From Eq.~(\ref{eq:timelikeksi4}) we can derive

\begin{equation}
\mathcal{E}^{MNPQ}\, \mathcal{J}^{[I}{}_{M} \tilde{\mathcal{J}}^{J}{}_{N}
\tilde{\mathcal{J}}^{K}{}_{P} \tilde{\mathcal{J}}^{L]}{}_{Q}=0\, ,  
\end{equation}

\noindent
whence the projections that would correspond to the hypers are automatically
satisfied.

From Eq.~(\ref{eq:ksi3}) we get 

\begin{eqnarray}
\label{eq:ksi3-1}
\langle\,\mathcal{E}^{a} \mid \mathcal{V}^{*\, i}\, \rangle
+\tfrac{1}{2\sqrt{2}}\delta^{a}{}_{0} \mathcal{E}^{iIJ}\frac{M_{IJ}}{|M|}
& = & 0\, ,\\
& & \nonumber \\
\label{eq:ksi3-2}
\langle\,\mathcal{E}^{a} \mid \mathcal{V}^{*\, i}\, \rangle  
M^{KI}
-{\textstyle\frac{1}{4}} \mathcal{E}^{i[I|J}V^{|K]}{}_{J}{}^{a} 
& = & 0\, .
\end{eqnarray}

\noindent
The first of these equations states first of all that 

\begin{equation}
\langle\,\mathcal{E}^{m} \mid \mathcal{V}^{*\, i}\, \rangle  =0\, ,
\end{equation}

\noindent
which, combined with Eqs.~(\ref{eq:otraksi}) implies by means of the
completeness relation Eq.~(\ref{eq:completeness2}) that

\begin{equation}
\mathcal{E}^{m}=0\, .  
\end{equation}

\noindent
Therefore, the only component of the Maxwell equations and Bianchi identities
that are not automatically satisfied due to supersymmetry, are
$\mathcal{E}^{0}$; secondly, for the projections onto equations of motion of
scalars in $N=2$ vector multiplets

\begin{equation}
\mathcal{E}^{i\,  KL}\mathcal{J}^{I}{}_{K}\mathcal{J}^{J}{}_{L}
=
-2\sqrt{2} \frac{M^{IJ}}{|M|} \langle\,\mathcal{E}^{a} \mid
\mathcal{V}^{*\, i}\, \rangle\, .
\end{equation}

\noindent
Contracting the second of these equations with $V_{a}|M|^{-2}$ we get 

\begin{equation}
\label{eq:timelikeksi5}
\langle\,\mathcal{E}^{a} \mid \mathcal{V}^{*\, i}\, \rangle  
\frac{M^{IJ}}{|M|}
-{\textstyle\frac{1}{2\sqrt{2}}}\delta^{a}{}_{0} 
\mathcal{E}^{iK[I}\mathcal{J}^{J]}{}_{K}
= 0\, ,
\end{equation}

\noindent
from which we get for the projections onto equations of motion of scalars in
$N=2$ hypermultiplets

\begin{equation}
\mathcal{E}^{i\,  KL}\mathcal{J}^{I}{}_{[K}\tilde{\mathcal{J}}^{J}{}_{L]}
=
0\, .
\end{equation}




For the special cases $N=5$ and $3$ we can define the $\mathrm{SU}(N)$ duals
of the scalar equations of motion:

\begin{equation}
\tilde{\mathcal{E}}_{I}\equiv {\textstyle\frac{1}{4!}}
\varepsilon_{IJKLM}\mathcal{E}^{JKLM}\, ,
\hspace{1cm}
\tilde{\mathcal{E}}^{i}{}_{I} \equiv  
{\textstyle\frac{1}{2}}
\varepsilon_{IJK}\mathcal{E}^{iJK}\, ,
\end{equation}

\noindent
and we can rewrite Eqs.~(\ref{eq:ksi4}) and (\ref{eq:ksi5}) in a more useful
form:

\begin{eqnarray}
\label{eq:ksi6}
\tilde{\mathcal{E}}_{I} \mathcal{J}^{I}{}_{J} & = & 0\, ,\\
& & \nonumber \\
\label{eq:ksi7}
\tilde{\mathcal{E}}^{i}{}_{I} \mathcal{J}^{I}{}_{J} & = & 0\, .   
\end{eqnarray}

Thus, in all cases the Einstein equations $\mathcal{E}^{0m},\mathcal{E}^{mn}$,
the Maxwell equations and Bianchi identities $\mathcal{E}^{m}$ and the scalar
equations $\mathcal{E}^{i\,
  KL}\mathcal{J}^{I}{}_{[K}\tilde{\mathcal{J}}^{J}{}_{L]}$ and
$\mathcal{E}^{MNPQ}\, \mathcal{J}^{[I}{}_{M} \tilde{\mathcal{J}}^{J}{}_{N}
\tilde{\mathcal{J}}^{K}{}_{P} \tilde{\mathcal{J}}^{L]}{}_{Q}$ are
automatically satisfied; the Einstein equation $\mathcal{E}^{00}$ and the
scalar equations $\mathcal{E}^{i\,
  KL}\mathcal{J}^{I}{}_{[K}\mathcal{J}^{J}{}_{L]}$ and $\mathcal{E}^{MNPQ}\,
\mathcal{J}^{[I}{}_{M} \mathcal{J}^{J}{}_{N} \tilde{\mathcal{J}}^{K}{}_{P}
\tilde{\mathcal{J}}^{L]}{}_{Q}$ are satisfied if the $0^{th}$ component of the
Maxwell equations and Bianchi identities $\mathcal{E}^{0}$ are satisfied. To
check that all the scalar equations of motion are, therefore, satisfied if
$\mathcal{E}^{0}$ are, it is convenient to make a detailed analysis case by
case.

\begin{description}
\item[$N=2$::] As mentioned before, Eq.~(\ref{eq:ksi3-1}) relates the
  complete scalar equations of motion to the $0^{th}$ component of the Maxwell
  equations an Bianchi identities. Therefore, we only need to solve
  $\mathcal{E}^{0}=0$.
\item[$N=3$::]  The KSIs Eqs.~(\ref{eq:timelikeksi5}) and (\ref{eq:ksi7})
can be combined into

\begin{eqnarray}
\label{eq:scalarKSIN=3}
\tilde{\mathcal{E}}^{i}{}_{I} 
& = & 
-2\sqrt{2}
\frac{\tilde{M}_{I}}{|M|}
\langle\,\mathcal{E}^{0} \mid \mathcal{V}^{*\, i}\, \rangle  
\, ,
\end{eqnarray}

\noindent
and we conclude that, as in the $N=2$ case, the only equation that needs to be
solved is $\mathcal{E}^{0}=0$.

\item[$N=4$::] As mentioned before, Eq.~(\ref{eq:timelikeksi4}) relates the
  complete scalar equation $\mathcal{E}^{IJKL}$ to $\mathcal{E}^{0}$ because
  in the $N=4$ case $\mathcal{E}^{IJKL} = \varepsilon^{IJKL}\mathcal{E}$,
  where $\mathcal{E}$ is the equation of motion of the complex scalar
  parametrizing $\mathrm{Sl}(2,\mathbb{R})/\mathrm{SO}(2)$. More explicitly,
  we have

\begin{equation}
\label{eq:scalarKSIN=4-1}
\mathcal{E} 
= 
-\sqrt{2}
\frac{\tilde{M}_{IJ}}{|\tilde{M}|}
\langle\,\mathcal{E}^{0} \mid \mathcal{V}^{*\, IJ}\, \rangle       
\, .
\end{equation}

\noindent
From Eq.~(\ref{eq:timelikeksi5}) and its $\mathrm{SU}(4)$ dual, using the
$N=4$ constraint $\mathcal{E}^{iIJ}=
\tfrac{1}{2}\varepsilon^{IJKL}\mathcal{E}_{iKL}$ we arrive at the
$N=4$-specific KSI

\begin{equation}
\label{eq:scalarKSIN=4-2}
\mathcal{E}_{iIJ}
= 
-2\sqrt{2}\left\{
\frac{\tilde{M}_{IJ}}{|\tilde{M}|}
\langle\,\mathcal{E}^{0} \mid \mathcal{V}^{*\, i}\, \rangle    
+
\frac{M_{IJ}}{|M|}
\langle\,\mathcal{E}^{0} \mid \mathcal{V}_{i}\, \rangle    
\right\}\, ,
\end{equation}

\noindent
which guarantees that, as in the foregoing cases, the matter scalar equations
of motion are satisfied if $\mathcal{E}^{0}=0$ is satisfied.

\item[$N=5$::] In this case we have to consider the $\mathrm{SU}(5)$ dual of
  Eqs.~(\ref{eq:timelikeksi4}) and (\ref{eq:ksi6}) which can be combined
  into the single identity

\begin{eqnarray}
\label{eq:scalarKSIN=5}
\tilde{\mathcal{E}}_{I} 
& = & 
-\sqrt{2}
\frac{\tilde{M}_{IJK}}{|M|}
\langle\,\mathcal{E}^{0} \mid \mathcal{V}^{*\, JK}\, \rangle  
\, ,
\end{eqnarray}

\noindent
which leads us to the same conclusion as in the previous cases.

\item[$N=6$::] In this case we have to consider the KSIs
  (\ref{eq:timelikeksi4}) involving $\mathcal{E}^{IJKL}$ and
  (\ref{eq:timelikeksi5}), involving $\mathcal{E}^{IJ}$ plus the constraint
  relating these equations of motion:
  $\mathcal{E}^{IJKL}=\tfrac{1}{2}\varepsilon^{IJKLMN}\mathcal{E}_{MN}$. Expressing
  both KSIs in terms of $\mathcal{E}^{IJ}$ only, we can combine them into

\begin{eqnarray}
\label{eq:scalarKSIN=6}
\mathcal{E}^{IJ}
& = & 
-2\sqrt{2} 
\frac{M^{IJ}}{|M|}
\langle\,\mathcal{E}^{0} \mid \mathcal{V}^{*}\, \rangle  
-\sqrt{2}
\frac{\tilde{M}^{IJKL}}{|M|}
\langle\,\mathcal{E}^{0} \mid \mathcal{V}_{KL}\, \rangle\, ,
\end{eqnarray}

\noindent
which brings us to the same conclusion as before.

\item[$N=8$::] The KSI (\ref{eq:timelikeksi4}) plus the constraint
  $\mathcal{E}^{IJKL}= \tfrac{1}{4!}\varepsilon^{IJKLMNPQ}\mathcal{E}_{MNPQ}$
  result in the KSI

\begin{equation}
\label{eq:scalarKSIN=8}
\mathcal{E}^{IJKL}
= 12\sqrt{2}
\left\{
\frac{M^{[IJ|}}{|M|}\langle\,\mathcal{E}^{0} \mid \mathcal{V}^{*\, |KL]}\, \rangle  
+\tfrac{1}{12}\frac{\tilde{M}^{IJKLMN}}{|M|}
\langle\,\mathcal{E}^{0} \mid \mathcal{V}_{MN}\, \rangle  
\right\}\, .
\end{equation}

\end{description}

In all cases the equations of motion of the scalars are automatically
satisfied if the $0^{th}$ component of the Maxwell equations and Bianchi
identities are. This will simplify the task of finding supersymmetric
solutions enormously as there is only one independent symplectic vector of
equations $\mathcal{E}^{0}$. On the other hand, to check consistency, we have
to check that all the supersymmetric configurations satisfy the above KSIs.







\section{$N\geq 2, d=4$ Killing Spinor Equations for the bilinears}
\label{sec-bilinearkses}


The supersymmetry rules in Sec.~(\ref{sec-kses}) induce differential relations
between the spinor-bilinears, defined in Section (\ref{sec-Fierz}), and the
various supergravity fields. As such, these relations contain the local
information of the supersymmetric configurations and the solutions and are
therefore the starting point in the deductive reconstruction process of the
supergravity fields from the KSEs. We start this process by enumerating said
differential relations.

From Eq.~(\ref{eq:kse1}) we get

\begin{eqnarray}
 \label{eq:dm}
\mathfrak{D}_{\mu}M_{IJ} 
-2iT_{K[I|}{}^{+}{}_{\mu\nu} V^{K}{}_{|J]}{}^{\nu}
& = & 0\, ,\\
& & \nonumber \\
\label{eq:DVIJ}
\mathfrak{D}_{\mu} V^{I}{}_{J\, \nu} 
+i\left\{ \left[M^{IK}T_{JK}{}^{+}{}_{\mu\nu} -\mathrm{h.c.}\right]
-\left[\Phi^{IK}{}_{(\mu|}{}^{\rho}T_{KJ}{}^{+}{}_{|\nu)\rho}
-\mathrm{h.c.}
\right]\right\} & = & 0\, . \label{eq:dv} 
\end{eqnarray}

From Eq.~(\ref{eq:kse2}) we get

\begin{eqnarray}
\label{eq:kse2-1}
 M^{KL}P_{KLIJ\mu} 
+6iT_{[IJ|}{}^{+}{}_{\mu\nu}V^{K}{}_{|K]}{}^{\nu} & = & 0\, ,\\
& & \nonumber \\
\label{eq:kse2-2}
P_{IJKL}\cdot V^{L}{}_{M} 
-{\textstyle\frac{3i}{2}}T_{[IJ}{}^{+}\cdot\Phi_{K]M} & = & 0\, .
\end{eqnarray}

From Eq.~(\ref{eq:kse4}) we get

\begin{eqnarray}
\label{eq:kse4-1}
M^{IJ}P_{iIJ\mu} +2iT_{i}{}^{+}{}_{\mu\nu}V^{\nu} & = & 0\, ,\\
& & \nonumber \\
\label{eq:kse4-2}
P_{iIJ}\cdot V^{J}{}_{K} -{\textstyle\frac{i}{2}}T_{i}{}^{+}\cdot
\Phi_{IK} & = & 0\, .
\end{eqnarray}

From Eq.~(\ref{eq:kse3}), which is only to be considered for $N=5$, we obtain

\begin{eqnarray}
\label{eq:kse3-1}
N=5 ::\hspace{2cm} P_{[IJKL}\cdot V^{N}{}_{M]} & = & 0\, ,\\
& & \nonumber \\
\label{eq:kse3-2}
N=5 ::\hspace{1.9cm} P_{[IJKL|\, \mu}M_{|M]N} & = & 0\, .
\end{eqnarray}

\noindent
The last equation can be written as

\begin{equation}
\label{eq:mattersusycondition3}
N=5 ::\hspace{2cm} \tilde{P}^{I}{}_{\mu}\ \mathcal{J}_{I}{}^{J}=0\, ,
\end{equation}

\noindent
where we have used the dual Vielbein
$\tilde{P}^{I}{}_{\mu}=\frac{1}{4!}\varepsilon^{IJKLM}P_{JKLM\, \mu}$.

As was said before, in the case of $N=3$ we must also take into account
Eq.~(\ref{eq:kse5}), which leads to

\begin{eqnarray}
\label{eq:kse5-1}
N=3 ::\hspace{2.2cm} P_{i[IJ}\cdot V^{L}{}_{K]}  & = &  0\, ,\\
& & \nonumber \\
\label{eq:kse5-2}
N=3 ::\hspace{2cm} P_{i[IJ|\, \mu}M_{|K]L} & = & 0\, .
\end{eqnarray}

\noindent
As in the $N=5$ case, we can use the dual Vielbein
$\tilde{P}^{iI}{}_{\mu}=\frac{1}{2}\varepsilon^{IJK}P_{iJK\, \mu}$ to rewrite
the last equations as

\begin{equation}
\label{eq:mattersusycondition4}
N=3 ::\hspace{2cm} \tilde{P}^{iI}{}_{\mu}\ \mathcal{J}_{I}{}^{J}=0\, .  
\end{equation}


\subsection{First consequences}
\label{sec-first}

Having enumerated the differential relations, we start the analysis by
expanding Eq.~(\ref{eq:kse2-1}), as to obtain

\begin{equation}
M^{KL}P_{KLIJ\mu} +2iT_{IJ}{}^{+}{}_{\mu\nu}V^{\nu} 
+4iT_{K[I|}{}^{+}{}_{\mu\nu}V^{K}{}_{|J]}{}^{\nu}
=0\, .
\end{equation}

Substituting Eq.~(\ref{eq:dm}) in the last term, we get

\begin{equation}
\label{eq:cij+}
C_{IJ}{}^{+}{}_{\mu}\equiv V^{\nu}T_{IJ}{}^{+}{}_{\nu\mu} =
-{\textstyle\frac{i}{2}} M^{KL}P_{KLIJ\mu} 
-i\mathfrak{D}_{\mu}M_{IJ}\, ,  
\end{equation}

\noindent
from which we can find $T_{IJ}{}^{+}$ by means of the following relation that
holds in the timelike case

\begin{equation}
\label{eq:tij+}
T_{IJ}{}^{+}  = V^{-2}[\hat{V}\wedge C_{IJ}{}^{+} 
+i\star (\hat{V}\wedge C_{IJ}{}^{+}) ]\, .
\end{equation}

Likewise from  Eq.~(\ref{eq:kse4-1}) we deduce

\begin{equation}
\label{eq:ti+}
C_{i}{}^{+}{}_{\mu} \equiv
V^{\nu}T_{i}{}^{+}{}_{\nu\mu} =
 -{\textstyle\frac{i}{2}}M^{IJ}P_{iIJ\mu}
\;\;
\longrightarrow\;\;
T_{i}{}^{+}  = V^{-2}[\hat{V}\wedge C_{i}{}^{+} 
+i\star (\hat{V}\wedge C_{i}{}^{+}) ]\, .
\end{equation}

\noindent
Eqs.~(\ref{eq:cij+},\ref{eq:ti+}) and (\ref{eq:dflij}) can then be used to 
find the complete field strengths, \textit{i.e.}

\begin{eqnarray}
C^{\Lambda +}{}_{\mu}\equiv V^{\nu}F^{\Lambda +}{}_{\nu\mu} 
& = & 
{\textstyle \frac{i}{2}}f^{*\Lambda IJ}C_{IJ}{}^{+}{}_{\mu} 
+i f^{*\, \Lambda i} C_{i}{}^{+}{}_{\mu} 
\nonumber \\
& & \nonumber \\
& = & 
{\textstyle\frac{1}{4}} M^{IJ}f^{*\Lambda KL}P_{IJKL\mu}
+{\textstyle\frac{1}{2}} M^{IJ}f^{*\, \Lambda i}P_{iIJ\mu} 
+{\textstyle\frac{1}{2}}f^{*\Lambda IJ}\mathfrak{D}_{\mu}M_{IJ} 
\nonumber \\
& & \nonumber \\
\label{eq:cl+}
& = & 
{\textstyle\frac{1}{2}}M^{IJ}\mathfrak{D}_{\mu}f^{\Lambda}{}_{IJ}
+{\textstyle\frac{1}{2}}f^{*\Lambda IJ}\mathfrak{D}_{\mu}M_{IJ}\, , 
\end{eqnarray}

\noindent
and

\begin{equation}
\label{eq:fl+}
F^{\Lambda +}  = V^{-2}[\hat{V}\wedge C^{\Lambda +} 
+i\star (\hat{V}\wedge C^{\Lambda +}) ]\, .
\end{equation}

The trace over $I$ over $J$ in Eq.~(\ref{eq:dv}) gives

\begin{equation}
\nabla_{\mu} V_{\nu} 
+i\left[M^{IJ}T_{IJ}{}^{+}{}_{\mu\nu} -\mathrm{c.c.}\right]=0\, ,  
\end{equation}

\noindent
which implies that $V^{\mu}$ is always a Killing vector 

\begin{equation}
\nabla_{(\mu}V_{\nu)}=0\, ,  
\end{equation}

\noindent
and that, had we been dealing with the null case ($M_{IJ}=0$), it would have
been covariantly constant.

Considering the equations involving the Vielbeine for each value of $N$, we
can derive the general result

\begin{equation}
\label{eq:timeindependentvielbeine}
V^{\mu}P_{IJKL\mu} = V^{\mu}P_{iIJ\mu} =0\, .
\end{equation}

The first of these equations together with the expression for
$T_{IJ}{}^{+}{}_{\mu\nu}V^{\nu}$, Eq.~(\ref{eq:tij+}), implies

\begin{equation}
\label{eq:timeindependentMIJ}
V^{\mu}\mathfrak{D}_{\mu}M_{IJ} =0\, .
\end{equation}

%
\subsection{Timelike case}
\label{sec-timelike}

We define the time coordinate $t$ by

\begin{equation}
V^{\mu}\partial_{\mu}\equiv\sqrt{2} \partial_{t}\, ,
\end{equation}

\noindent
which implies that all the fields are (covariantly) time-independent.  Taking
into account that $V^{2}=2|M|^{2}$ and the above choice of coordinate,
$\hat{V}$ must take the form

\begin{equation}
\label{eq:hatV}
\hat{V} \equiv V_{\mu}dx^{\mu}= \sqrt{2}|M|^{2}(dt+\omega)\, 
\end{equation}

\noindent
where $\omega=\omega_{\underline{m}}dx^{m}$ is a time-independent 1-form to be
determined.  We can use the 1-form $\hat{V}$ to construct the $0^{th}$ component of
a Vielbein basis $\{e^{a}\}$

\begin{equation}
e^{0}\equiv \tfrac{1}{\sqrt{2}}|M|^{-1}\hat{V}\, .  
\end{equation}

\noindent
The other three 1-forms of the basis $\{e^{1},e^{2},e^{3}\}$ will be chosen
arbitrarily\footnote{It is worth stressing the differences with the procedure
  followed in the $N=2$ case in Ref.~\cite{Huebscher:2006mr}: in the $N=2$
  case one can use the well-known constant Pauli matrices and construct
  $\{e^{1},e^{2},e^{3}\}$ decomposing the vector bilinear $V^{I}{}_{J\, \mu}$
  with respect to $\{\sigma^{1},\sigma^{2},\sigma^{3}\}$. In the general case
  there are {\it a priori} no constant $N\times N$ Pauli matrices available
  and we are forced to choose $\{e^{1},e^{2},e^{3}\}$ first, and then use them
  to construct the $N\times N$ Pauli matrices, which generically will be
  non-constant: see Appendix~\ref{sec-Fierz} for more detail.}.  In general
none of the remaining vector bilinears is an exact 1-form: with the
available information we can only say that the 4-dimensional metric takes the
form

\begin{equation}
\label {eq:timelikemetric}
ds^{2} \; =\; |M|^{2} (dt+\omega)^{2} 
-|M|^{-2}\gamma_{\underline{m}\underline{n}}dx^{m}dx^{n}\, ,
\end{equation}

\noindent
where the 3-dimensional metric $\gamma_{\underline{m}\underline{n}}$ also has
to be determined. The 1-forms $\hat{V}^{m}$ defined in Eq.~(\ref{eq:1formsVm})
can be taken as Dreibeine for the metric
$\gamma_{\underline{m}\underline{n}}$. We are going to derive from
Eq.~(\ref{eq:DVIJ}), which contains a great deal of information, equations for
$\hat{V}$, $\hat{V}^{m}$ and the matrices $(\sigma^{m})^{I}{}_{J}$, defined in
Eq.~(\ref{eq:sigmamdef}), that will determine $\omega$ and
$\gamma_{\underline{m}\underline{n}}$.

Using the decompositions (\ref{eq:decompositionVIJ},\ref{eq:p2}) and the
expression for the graviphotons field strengths, Eq.~(\ref{eq:tij+}), in
Eq.~(\ref{eq:DVIJ}) we get

\begin{eqnarray}
\label{eq:dV}
d\hat{V} + |M|^{-2}
\left\{  \hat{V} \wedge d|M|^{2} 
+i\star \left[  \hat{V} \wedge (M^{IJ}\mathfrak{D}M_{IJ}
-M_{IJ}\mathfrak{D}M^{IJ} ) \right]
\right\} & = & 0\, ,\\
& & \nonumber \\   
\label{eq:dVx}
d\hat{V}^{m}  
+\tfrac{1}{2}\mathrm{Tr}\, 
\left(
\sigma^{m}\mathfrak{D}\sigma^{n}
\right)
\wedge \hat{V}^{n}
& = & 0\, ,\\
& & \nonumber \\   
\label{eq:Dmsnsym}
\mathfrak{D}_{m}\sigma^{n}  +\mathfrak{D}_{n}\sigma^{m} & = & 0\, ,\\  
& & \nonumber \\   
\label{eq:Dmsnantisym}
\varepsilon_{mnp}\left[\mathfrak{D}_{n}\sigma^{p} 
+\tfrac{1}{2}\mathrm{Tr}\left(\sigma^{p}\mathfrak{D}_{n}\sigma^{q} \right)
\sigma^{q} \right]
-i\left(\mathfrak{D}_{m}\mathcal{J}\mathcal{J}
-\mathcal{J}\mathfrak{D}_{m}\mathcal{J}\right)
 & = & 0\, ,\\  
& & \nonumber \\   
\label{eq:DJ}
\mathfrak{D}_{m}\mathcal{J}^{I}{}_{J} 
+2i|M|^{-2}\varepsilon_{mnp}
\left[\mathfrak{D}_{n}M_{JK}(\sigma^{p})^{K}{}_{L}M^{LI} -\mathrm{h.c.} 
\right] & = & 0\, .
\end{eqnarray}

\noindent
Observe that, even though the $\sigma$-matrices bear indices $m,n$ and $p$,
these indices are not tangent space indices and the covariant derivatives
acting on them is the $\mathrm{U}(N)$ connection $\Omega$ only.

If we act with $\mathcal{J}^{I}{}_{L}$ on Eq.~(\ref{eq:dm}) and use the
expression for the graviphoton field strengths Eq.~(\ref{eq:tij+}) and the
trace of Eq.~(\ref{eq:Dmsnsym}), we get
$\mathcal{J}\mathfrak{D}\mathcal{J}=0$, which together with its Hermitean
conjugate imply the very important condition

\begin{equation}
\label{eq:DJ=0}
\mathfrak{D}\mathcal{J}\ =\ 0\, .
\end{equation}

This equation does not imply that it is possible to choose a gauge in which
$d\mathcal{J}=0$ because the theories we are considering are only invariant
under global $\mathrm{U}(N)$ transformations and not under \textit{arbitrary}
gauge transformations (the connection $\Omega$ is a composite field).
Nevertheless, observe that $\mathcal{J}$ is constant in the $\mathrm{U}(2)$
directions of the Killing spinors:

\begin{equation}
\label{eq:JdJJ=0}
\mathcal{J} d\mathcal{J} \mathcal{J}=0\, ,  
\end{equation}

\noindent
as follows from its idempotency $\mathcal{J}^{2} =\mathcal{J}$.  On the other
hand, this condition will allow us to relate consistently each supersymmetric
configuration to a truncation to an $N=2$ theory with vector supermultiplets
and hypermultiplets: $\mathcal{J}$ projects the $\mathrm{U}(N)$ space onto an
$\mathrm{U}(2)$ subspace, which defines the associated $N=2$ truncation. Using
$\mathcal{J}$ we are going to be able to project the scalar Vielbeine
$P_{IJKL}$ and $P_{i\, IJ}$ onto scalar Vielbeine belonging to the vector
supermultiplets or the hypermultiplets of the truncation.

The integrability condition of $\mathfrak{D}\mathcal{J}=0$ is

\begin{equation}
[R(\Omega),\mathcal{J}] =0\, , 
\end{equation}

\noindent
which restricts the holonomy of the pullback of the connection of the scalar
manifold to the group generated by the $\mathrm{U}(N)$ subalgebra that
commutes with $\mathcal{J}$; this group is $\mathrm{U}(2)\otimes
\mathrm{U}(N-2)$, the first factor being generated by
$\{\mathcal{J},\sigma^{1},\sigma^{2},\sigma^{3}\}$.

Since $R(\Omega)$ can be expressed in terms of the scalar Vielbeine using
Eq.~(\ref{eq:RO=PP}), the above condition is a condition on the
Vielbeine. Below, we are going to derive several conditions for the Vielbeine
that will ensure that the above condition is satisfied.

Another important consequence of the condition $\mathfrak{D}\mathcal{J}=0$ is

\begin{equation}
\label{eq:DMM}
\mathfrak{D}M^{IJ} = |M|^{-2}M^{IJ}M_{KL}\mathfrak{D}M^{KL}\, ,  
\end{equation}

\noindent
which leads to relations such as

\begin{equation}
\label{eq:DMM=0}
\mathfrak{D}M^{[IJ}\mathfrak{D}M^{K]L} = 0\, ,
\end{equation}

\noindent
and solves Eq.~(\ref{eq:DJ}).

Let us continue by analyzing Eq.~(\ref{eq:dV}): taking the exterior derivative
of $\hat{V}$ in Eq.~(\ref{eq:hatV}) and comparing it with Eq.~(\ref{eq:dV}) we
find that

\begin{equation}
\label{eq:do}
d\omega =\frac{i}{\sqrt{2} |M|^{4}}\star\left[(M^{IJ}\mathfrak{D}M_{IJ}
-M_{IJ}\mathfrak{D}M^{IJ} ) \wedge \hat{V} \right]\, ,  
\end{equation}

\noindent
which can be rewritten as an equation in the background of the 3-dimensional
spatial metric:

\begin{equation}
\label{eq:do23-d}
(d\omega)_{mn} =-\frac{i}{|M|^{4}}\varepsilon_{mnp}
(M^{IJ}\mathfrak{D}_{p}M_{IJ}
-M_{IJ}\mathfrak{D}_{p}M^{IJ})\, .  
\end{equation}

\noindent
Using the symplectic vectors $\mathcal{R}$ and $\mathcal{I}$ defined in
Eq.~(\ref{eq:RandIdef}) and the constraint $M^{[IJ}M^{K]L}=0$,
Eq.~(\ref{eq:mm3}), we find that

\begin{equation}
M^{IJ}\mathfrak{D}_{m}M_{IJ}
-M_{IJ}\mathfrak{D}_{m}M^{IJ} = 2i|M|^{4}\langle\,\mathcal{I}\mid
\partial_{m}\mathcal{I}\rangle\, ,  
\end{equation}

\noindent
and then we can rewrite the equation for $\omega$ in terms of $\mathcal{I}$

\begin{equation}
\label{eq:oidi}
(d\omega)_{mn} =  2\epsilon_{mnp}
\langle\,\mathcal{I}\mid \partial^{p}\mathcal{I}\, \rangle\, ,
\end{equation}

\noindent
and $|M|$ in terms of $\mathcal{R}$ and $\mathcal{I}$

\begin{equation}
\label{eq:MRI}
|M|^{-2}=  \langle\,\mathcal{R}\mid \mathcal{I}\, \rangle\, ,
\end{equation}

\noindent
which are identical to the ones obtained in
Refs.~\cite{Meessen:2006tu,Bellorin:2006xr} for $N=2$ theories coupled to
vector multiplets and with the same integrability condition, namely

\begin{equation}
\label{eq:oidiintcond}
\langle\,\mathcal{I}\mid \nabla_{(3)}^{2}\mathcal{I}\, \rangle =0\, .
\end{equation}

Let us now move on to Eq.~(\ref{eq:dVx}): it can be interpreted as Cartan's
first structure equation for a torsionless connection
$\varpi^{mn}=-\varpi^{nm}$ on the 3-dimensional space

\begin{equation}
d\hat{V}^{m}  
-\varpi^{mn}\wedge \hat{V}^{n}=0\, ,
\end{equation}

\noindent
where the connection can be read off and is

\begin{equation}
\label{eq:3dconnection}
\varpi^{mn}=  -\tfrac{1}{2}\mathrm{Tr}\, [\sigma^{m}\mathfrak{D}\sigma^{n}]
= i\varepsilon^{mnp}\mathrm{Tr}\, [\sigma^{p}\Omega] 
-\tfrac{1}{2}\mathrm{Tr}\, [\sigma^{m}d\sigma^{n}]\, .
\end{equation}

This equation relates the spin connection of the 3-dimensional transverse
space to the pullback of the connection of the scalar manifold. This spin
connection is constrained by Eq.~(\ref{eq:Dmsnsym}): multiplying by
$\sigma^{p}$ and taking the trace, we find that  

\begin{equation}
  \varpi_{(mn)p} =0\, ,\,\,\, \Rightarrow\,\,\, \varpi_{mnp}=\varpi_{[mnp]}\, ,
\end{equation}

\noindent
which is a gauge condition associated to our choices.

Defining a new covariant derivative $\hat{\mathfrak{D}}=\mathfrak{D}+\varpi$,
where $\varpi^{mn}$ acts on the upper $m,n$ indices of the $\sigma$
matrices\footnote{Explicitly, $\hat{\mathfrak{D}}_{m}\sigma^{n}\equiv
  \mathfrak{D}_{m}\sigma^{n} -\varpi_{m}{}^{np}\sigma^{p}$. We do not
  distinguish between upper and lower flat 3-dimensional indices.} we can
rewrite now Eqs.~(\ref{eq:Dmsnsym}) and (\ref{eq:Dmsnantisym}) in the combined
form

\begin{eqnarray}
\label{eq:ds=0}
\hat{\mathfrak{D}}_{m}\sigma^{n}  = 
0
\, .
\end{eqnarray}


\noindent
The integrability condition of this equation relates the curvature 2-form of
$\varpi^{mn}$ to an $\mathfrak{su}(2)$ projection the curvature of the
pullback of the connection of the scalar manifold $\Omega$:

\begin{equation}
\label{eq:Rw=RO}
R^{mn}(\varpi) =i\varepsilon^{mnp}\, \mathrm{Tr}\,[\sigma^{p}R(\Omega)]\, . 
\end{equation}

\noindent
If we compute the curvature $R^{mn}(\varpi)$ using Eq.~(\ref{eq:3dconnection})
we find on the r.h.s.~the extra term 

\begin{equation}
i\varepsilon^{mnp}\, 
\mathrm{Tr}\,[\mathcal{J}d\sigma^{p}\mathcal{J} \wedge \Omega]\, ,
\end{equation}

\noindent 
which must vanish for consistency. We are going to impose the condition

\begin{equation}
\label{eq:JdsJ=0}
\mathcal{J}d\sigma^{p}\mathcal{J} =0\, ,  
\end{equation}

\noindent
which says that the $\sigma^{m}$ matrices are constant in the $\mathrm{U}(2)$
directions of the Killing spinors, just as $\mathcal{J}$.  We have not found a
better proof of this condition, but we shall see that it is the simplest
condition that solves the KSEs.

Using Eq.~(\ref{eq:RO=PP}) we can rewrite Eq.~(\ref{eq:Rw=RO}) in a form that
can be compared directly with the $\mathrm{SU}(2)$ curvature and quaternionic
structures of the quaternionic-K\"ahler manifold in which the scalars of $N=2$
hypermultiplets live. Then Eq.~(\ref{eq:Rw=RO}) relates the curvature of the
spatial 3-dimensional metric $\gamma$ with the $\mathrm{SU}(2)$ curvature of
the hyperscalars, completely analogous to what happens in the $N=2$ case with
hypermultiplets \cite{Huebscher:2006mr}. To find the projections of the scalar
Vielbeine that correspond to the hyperscalars in the associated $N=2$
truncation defined by $\mathcal{J}$, we first use Eqs.~(\ref{eq:Rw=RO}) and
(\ref{eq:RO=PP}) to write the Ricci tensor of $\gamma$ as

\begin{equation}
R(\gamma)_{mn} = -\frac{i}{N-2}\varepsilon^{npq}(\sigma^{q})^{I}{}_{J}
[P^{*\, JKLM}{}_{[m|}P_{IKLM\, |p]} +2 P^{*\, i\,JK}{}_{[m|}P_{i\, IK\, |p]}
]\, .  
\end{equation}

{}Further identities are needed: using the decompositions
(\ref{eq:decompositionVIJ},\ref{eq:p2}) and the time-independence of the
scalars Eq.~(\ref{eq:timeindependentvielbeine}) in Eqs.~(\ref{eq:kse2-2}) and
(\ref{eq:kse4-2}), together with the expressions for the supergravity and
matter vector field strengths Eqs.~(\ref{eq:cij+}-\ref{eq:ti+}), we get the
following constraints on the scalar Vielbeine:

\begin{eqnarray}
\left[ 
P_{IJKL\, m}
-3 |M|^{-2}M^{PQ}P_{PQ[IJ|\, m} M_{|K] L} 
\right]
(\sigma^{m})^{L}{}_{M} 
& = & 0\, ,\\
& & \nonumber \\  
P_{i\, MN\, m} \left(\delta^{MN}{}_{IJ} 
-\mathcal{J}^{M}{}_{[I}\mathcal{J}^{N}{}_{J]} \right) 
(\sigma^{m})^{J}{}_{K} 
& = & 0\, ,
\end{eqnarray}

\noindent
which can be rewritten in the form\footnote{These equations should be compared
  with the conditions that supersymmetry imposes on the pullbacks of the
  quaternionic Vielbeine in $N=2$ theories \cite{Huebscher:2006mr}.}

\begin{eqnarray}
\label{eq:PIJKLcondition}
P_{IJKL\, m}\, \mathcal{J}^{I}{}_{[M}
\tilde{\mathcal{J}}^{J}{}_{N} \tilde{\mathcal{J}}^{K}{}_{P}
\tilde{\mathcal{J}}^{L}{}_{Q]} (\sigma^{m})^{Q}{}_{R} 
& = & 0\, ,\\
& & \nonumber \\  
\label{eq:PiIJcondition}
P_{i\, IJ\, m}\, \mathcal{J}^{I}{}_{[K}
\tilde{\mathcal{J}}^{J}{}_{L]}(\sigma^{m})^{L}{}_{M} 
& = & 0\, .
\end{eqnarray}

Using them in the above equation, the Ricci tensor of $\gamma$ takes the
form\footnote{For $N=2$ the r.h.s.~vanishes identically, as the formalism used
  only takes into account vector multiplets.}

\begin{equation}
\label{eq:Rg}
  \begin{array}{rcl}
R(\gamma)_{mn} &  = & 
-{\displaystyle\frac{1}{N-2}}
\left[P_{IJKL\, (m|}
\mathcal{J}^{I}{}_{M}
\tilde{\mathcal{J}}^{J}{}_{N}
\tilde{\mathcal{J}}^{K}{}_{P}
\tilde{\mathcal{J}}^{L}{}_{Q}
P^{*\, MNPQ}{}_{|n)} 
\right.
\\
& & \\
& & 
\left.
+2 P_{i\,IJ\, (m|}\mathcal{J}^{I}{}_{M}
\tilde{\mathcal{J}}^{J}{}_{N}
P^{*\, i\, MN}{}_{|n)}
\right]\, .
\end{array}
\end{equation}

The hyperscalar Vielbeine in the associated $N=2$ truncation are clearly
identified in this expression. The conditions for a flat 3-dimensional metric,
or said differently the \textit{no-hypers} conditions, are therefore

\begin{eqnarray}
\label{eq:no-hyperssugra}
P_{IJKL}\, 
\mathcal{J}^{I}{}_{[M}
\tilde{\mathcal{J}}^{J}{}_{N}
\tilde{\mathcal{J}}^{K}{}_{P}
\tilde{\mathcal{J}}^{L}{}_{Q]}   & = & 0\, ,\\
& & \nonumber \\
\label{eq:no-hypersmatter}
P_{i\,IJ}\, 
\mathcal{J}^{I}{}_{[M}
\tilde{\mathcal{J}}^{J}{}_{N]}
& = & 0\, .
\end{eqnarray}


\section{Solving the KSEs}
\label{sec-timelikekses}

We have thus far obtained the following necessary conditions for a field configuration to 
admit at least one Killing spinor and to lie in the timelike class of solutions:

\begin{enumerate}
\item All the fields are time-independent and related to a complex,
  antisymmetric matrix $M^{IJ}$ satisfying $M^{[IJ}M^{K]L}=0$, from which we
  must construct the covariantly constant projection $\mathcal{J}^{I}{}_{J}$,
  and to generalized Pauli matrices $(\sigma^{m})^{I}{}_{J}$ which
  must satisfy Eqs.~(\ref{eq:sigmaprop1}-\ref{eq:sigmaprop8}) and
  (\ref{eq:JdsJ=0}).

\item The scalars have to satisfy Eqs.~(\ref{eq:PIJKLcondition}) and
  (\ref{eq:PiIJcondition}); in the special cases of  $N=3$ and $5$ they further need to satisfy
  Eqs.~(\ref{eq:mattersusycondition3}) and (\ref{eq:mattersusycondition4}).

\item The vector field strengths are given in terms of the scalars and the
  matrix $M^{IJ}$ by Eqs.~(\ref{eq:cij+}-\ref{eq:ti+})\footnote{Simpler
    expressions for the vector field strengths will be given in the next section.}.

\item The spacetime metric is of conforma-stationary form,
  Eq.~(\ref{eq:timelikemetric}), where
  \begin{enumerate}
  \item The 1-form $\omega$ is related to the matrix $M^{IJ}$, the scalar
  fields (through the pullback of the scalar connection) and the 3-dimensional
  transverse metric $\gamma_{mn}$ through Eq.~(\ref{eq:oidi}).

\item The 3-dimensional metric is related to the scalars and the generalized
  Pauli matrices by Eq.~(\ref{eq:3dconnection}) which relates its spin
  connection to an $\mathrm{SU}(2)$ projection of the pullback of the
  connection of the scalar manifold.

  \end{enumerate}

\end{enumerate}

We are going to see that these necessary conditions are also sufficient: let
us start by plugging our result for $T_{i}$ Eq.~(\ref{eq:ti+}) into
Eq.~(\ref{eq:kse4}), leading to

\begin{equation}
P_{iKL\, m}\gamma^{m}
\left[
\delta_{I}{}^{K}\epsilon^{L}
-\tfrac{i}{\sqrt{2}}|M|^{-1}M^{KL}\gamma^{0}\epsilon_{I}
\right]=0\, .
\end{equation}

\noindent
Decomposing now

\begin{equation}
P_{iKL\, m} = P_{iMN\, m}\mathcal{J}^{M}{}_{[K}\mathcal{J}^{N}{}_{L]}
+P_{iMN\, m}(\delta^{MN}{}_{KL}-\mathcal{J}^{M}{}_{[K}\mathcal{J}^{N}{}_{L]})
\, ,  
\end{equation}

\noindent
we get

\begin{equation}
P_{iMN\, m}|M|^{-1}M^{MN}
\gamma^{m}\left[
|M|^{-1}M_{IL}\epsilon^{L}
-\tfrac{i}{\sqrt{2}}\gamma^{0}\epsilon_{I}
\right]
+P_{iMN\, m}
(\delta^{MN}{}_{KL}-\mathcal{J}^{M}{}_{[K}\mathcal{J}^{N}{}_{L]})
\gamma^{m}\epsilon^{L}
=0\, .  
\end{equation}

\noindent
Each of the two terms has to vanish separately because they depend on
independent components of $P_{iIJ\, m}$. The first term can vanish in two
different ways:

\begin{enumerate}

\item $P_{iMN\, m}M^{MN}=0$ (vanishing matter vector field strengths $T_{i}$
  (\ref{eq:ti+})). In this case, the generic way to make the second term to
  vanish is to impose\footnote{Compare this equation with Eq.~(4.35) of
    Ref.~\cite{Huebscher:2006mr}.}

\begin{equation}
\label{eq:hyperprojector}
\Pi^{m\pm\, I}{}_{J}\epsilon^{J} \equiv 
\tfrac{1}{2}[\delta^{I}{}_{J} \pm \gamma^{0(m)}(\sigma^{(m)})^{I}{}_{J}]\epsilon^{J}=0\, ,
\end{equation}

\noindent
for each value of $m$ for which $P_{iIJ\, m}\neq 0$ and then use
Eq.~(\ref{eq:PiIJcondition}).  The consistency of this condition for a given
$m$ requires\footnote{These projectors satisfy $(\Pi^{m\pm})^{2} =\Pi^{m\pm}
  -\tfrac{1}{4}(1-\mathcal{J})$ and $[\Pi^{m\pm},\Pi^{n\pm}]=0$.}

\begin{equation}
\label{eq:1-J}
(\delta^{I}{}_{J}-\mathcal{J}^{I}{}_{J})\epsilon^{J}=0\, ,  
\end{equation}

\noindent
which reduces the number of unbroken supersymmetries to just two ({\em i.e.\/} eight real
independent supercharges), out of which only one half (i.e.~$1/N$) survives the
projection Eq.~(\ref{eq:hyperprojector}) for one given value of $m$. If we
have to impose another projector of the same kind, the number of unbroken
supersymmetries is lowered by another factor of $1/2$. In the generic case we
will have to impose all three projectors and the supersymmetry preserved is
just one (i.e.~$1/(4N)$ of the total).

If Eq.~(\ref{eq:1-J}) is satisfied and $P_{iMN\, m}
(\delta^{MN}{}_{KL}-\mathcal{J}^{M}{}_{[K}\mathcal{J}^{N}{}_{L]})
\mathcal{J}^{L}{}_{J}=0$ (which is identical to the ``no-hypers'' condition
Eq.~(\ref{eq:no-hypersmatter}), we do not need to impose
Eq.~(\ref{eq:hyperprojector}), which is associated to the hypermultiplets in
the associated $N=2$ truncation. It is clear that the projected scalar
Vielbeine $P_{iMN\, m}\mathcal{J}^{M}{}_{[K}\mathcal{J}^{N}{}_{L]}$ correspond
to the complex scalar of the vector multiplets of the $N=2$ truncation.

\item If $P_{iMN\, m}M^{MN}\neq 0$ then we have to impose

\begin{equation}
\label{eq:genericsusyprojector}
\epsilon_{I} 
+i\sqrt{2}|M|^{-1}M_{IJ}\gamma^{0}\epsilon^{J}
= 0\, ,  
\end{equation}

\noindent
which is consistent only if Eq.~(\ref{eq:1-J}) is satisfied, which means that,
generically, $1/(2N)$ of the total amount of available supercharges are preserved by this
condition.

The second term vanishes when we impose again the generic condition
Eq.~(\ref{eq:hyperprojector}), which is compatible with
Eq.~(\ref{eq:genericsusyprojector}), and use Eq.~(\ref{eq:PiIJcondition}).
Again, if Eq.~(\ref{eq:no-hypersmatter}) is satisfied, the condition
Eq.~(\ref{eq:hyperprojector}) is unnecessary.

\end{enumerate}

In the case of $N=3$ supergravity we have to consider the KSE
Eq.~(\ref{eq:kse5}), which is readily seen to be solved by the condition
Eq.~(\ref{eq:mattersusycondition3}). Observe that this condition automatically
implies the ``no-hypers'' condition, in agreement with the absence of
hypermultiplets in the truncations from $N=3$ to $N=2$. Therefore, in $N=3$
supergravity the only projector that ever needs to be imposed on the Killing
spinors is Eq.~(\ref{eq:genericsusyprojector}).

Let us then consider the KSE Eq.~(\ref{eq:kse2}). Substituting our result for
$T_{IJ}$, Eqs.~(\ref{eq:cij+}) and (\ref{eq:tij+}), we can immediately write
it as

\begin{equation}
  \begin{array}{rcl}
\left[P_{IJKM\, m} -3 \left(|M|^{-2}M^{MN}P_{MN[IJ|\, m}M_{|K]L}  
+2|M|^{-2} \mathfrak{D}_{m}M_{[IJ}M_{K]L}\right)\right]\gamma^{m}\epsilon^{L}
& & \\
& & \\
+3\left(|M|^{-2}M^{MN}P_{MN[IJ|\, m}  
+2|M|^{-2} \mathfrak{D}_{m}M_{[IJ|}\right)\gamma^{m}
\left(
|M|^{-1}M_{|K]L}\epsilon^{L} -\tfrac{i}{\sqrt{2}}\gamma^{0}\epsilon_{|K]}
\right)
& = & 0\, .
\end{array}
\end{equation}

Again, we can distinguish two different cases:

\begin{enumerate}

\item $M^{MN}P_{MNIJ\, m} +2\mathfrak{D}_{m}M_{IJ}=0$, which implies the
  vanishing of the vector field strengths (\ref{eq:cij+}) in the graviton
  supermultiplet.  In this case, the equation can generically be solved by
  imposing the projector Eq.~(\ref{eq:hyperprojector}) on the Killing spinors
  and using the constraint Eq.~(\ref{eq:PIJKLcondition}). If $P_{IJKM\,
    m}\mathcal{J}^{M}{}_{L}=0$, equivalent in this case to the ``no-hypers''
  condition Eq.~(\ref{eq:no-hyperssugra}), then the condition
  Eq.~(\ref{eq:1-J}) suffices.

\item $M^{MN}P_{MNIJ\, m} +2\mathfrak{D}_{m}M_{IJ} \neq 0$: in this case we
  need to impose the projectors Eq.~(\ref{eq:genericsusyprojector}) and, to
  cancel the first term we have to impose Eq.~(\ref{eq:hyperprojector}) unless
  $P_{IJKL\, m}$ satisfies\footnote{Here we have used Eq.~(\ref{eq:DMM=0}) to
    simplify the expression.}

\end{enumerate}

\begin{equation}
\label{eq:mattersusycondition2}
P_{IJKM\, m}\mathcal{J}^{M}{}_{L} -3 |M|^{-2}M^{MN}P_{MN[IJ|\, m}M_{K]L}=0\, ,
\end{equation}

\noindent
which implies the ``no-hypers'' condition Eq.~(\ref{eq:no-hyperssugra}).

For $N=5$ we also have to consider the KSE Eq.~(\ref{eq:kse3}): this equation
is immediately solved by the condition Eq.~(\ref{eq:kse3-2}), or equivalently
(\ref{eq:mattersusycondition3}), which is a particular instance of
Eq.~(\ref{eq:mattersusycondition2}) implying once again the ``no-hypers''
condition (\ref{eq:no-hyperssugra}). Therefore, in the $N=5$ case we only need
to impose the projection Eq.~(\ref{eq:genericsusyprojector}).

Using the supersymmetry conditions that we have used to solve the previous
KSEs plus $\mathfrak{D}\mathcal{J}=0$, it is easy to see that the $0^{th}$
component of the KSE Eq.~(\ref{eq:kse1}) is satisfied, while the $m^{th}$
component reduces to the equation in 3-dimensional transverse space

\begin{equation}
\label{eq:reducedKSE1}
\mathfrak{D}_{m} \epsilon_{I} 
-|M|^{-2}\mathfrak{D}_{m}M_{IK}M^{JK} \epsilon_{J}=0\, ,
\end{equation}

\noindent
where 

\begin{equation}
  \begin{array}{rcl}
\mathfrak{D}_{m}
\epsilon_{I} 
& = & 
(\partial_{m}+\tfrac{1}{4}\varpi_{mnp}\gamma^{np})\epsilon_{I}
-\Omega^{J}{}_{I}\epsilon_{J}
= 
\partial_{m}\epsilon_{I} +\left[\pm\tfrac{i}{4}\varpi_{mnp}\varepsilon^{npq}
(\sigma^{q})^{J}{}_{I}-\Omega^{J}{}_{I}\right]\epsilon_{J}\, ,\\ 
\end{array}
\end{equation}

\noindent
upon use of the condition Eq.~(\ref{eq:hyperprojector})\footnote{Acting on
  this equation with the projector $\tilde{\mathcal{J}}^{I}{}_{L}$ we find the
  integrability condition $\mathfrak{D}\mathcal{J}=0$.}.

\noindent
From Eqs.~(\ref{eq:3dconnection}) and (\ref{eq:JdsJ=0}) we obtain

\begin{equation}
\label{eq:esaconaxion}
\pm\tfrac{i}{4}\varpi_{mnp}\varepsilon^{npq} \sigma^{q}  
=
\mp \mathcal{J}\Omega \mathcal{J} \pm \tfrac{1}{2} \mathrm{Tr}\,
[\mathcal{J}\Omega]\, , 
\end{equation}

\noindent
and from $\mathfrak{D}\mathcal{J}=0$ we get 

\begin{equation}
 \mathcal{J}\Omega \tilde{\mathcal{J}}=\mathcal{J}d\mathcal{J}
=\tfrac{1}{4}(\mathcal{J}d\mathcal{J}+\sigma^{m}d\sigma^{m})\, .
\end{equation}

The second term in Eq.~(\ref{eq:reducedKSE1}) can be put in the form

\begin{equation}
  \begin{array}{rcl}
|M|^{-2}\mathfrak{D}M_{IK}M^{JK} \epsilon_{J}
&  = &
\tfrac{1}{2}
\left[
\mathfrak{D}\mathcal{J}^{J}{}_{I} 
+|M|^{-2}\mathfrak{D}M_{MN}M^{MN}\mathcal{J}^{J}{}_{I}
\right] \epsilon_{J}\\
& & \\
& = & 
\tfrac{1}{2}
\left[2i\xi +\tfrac{1}{2}|M|^{-2}\partial|M|^{2}
-\mathrm{Tr}(\mathcal{J}\Omega)\right]
 \epsilon_{J}\, ,
\end{array}
\end{equation}

\noindent
where

\begin{equation}
\label{eq:xi}
\xi \equiv \tfrac{i}{4}|M|^{-2}(dM^{MN}M_{MN} -dM_{MN}M^{MN})\, .  
\end{equation}

Putting all this information together and choosing the upper sign so the terms
$\mathrm{Tr}(\mathcal{J}\Omega_{m})$ cancel, we can rewrite the reduced KSE
using 3-dimensional differential forms as

\begin{equation}
d\hat{\epsilon} 
-\hat{\epsilon}[i\xi 
+\tfrac{1}{4}(\mathcal{J}d\mathcal{J}+\sigma^{m}d\sigma^{m})]=0\, ,
\end{equation}

\noindent
where we have defined the $\mathrm{U}(N)$ row vector $\hat{\epsilon}_{I}\equiv
|M|^{-1/2}\epsilon_{I}$. The integrability condition of this equation 

\begin{equation}
\mathcal{J}[id\xi +\tfrac{1}{4}(d\mathcal{J}\wedge
d\mathcal{J}+d\sigma^{m}\wedge d\sigma^{m})]=0\, ,  
\end{equation}

\noindent
is identically satisfied\footnote{Here and in Eq.~(\ref{eq:esaconaxion}) we
  have used $\mathcal{J}d\sigma^{m}\mathcal{J}=0$.}.

This shows that the necessary conditions for supersymmetry enumerated at the
beginning of this section are also sufficient. Furthermore, we have shown that
the Killing spinors generically satisfy the condition Eq.~(\ref{eq:1-J}),
which preserves $2/N$ supersymmetries; if the supergravity or matter vector
field strengths are non-vanishing, then they also satisfy the condition
Eq.~(\ref{eq:genericsusyprojector}), which breaks a further $1/2$ of the
supersymmetries and, if one of the scalar Vielbein projections $P_{IJKL\,
  m}\mathcal{J}^{I}{}_{M}\tilde{\mathcal{J}}^{J}{}_{N}
\tilde{\mathcal{J}}^{K}{}_{P} \tilde{\mathcal{J}}^{L}{}_{Q}$ or $P_{i\,IJ\,
  m}\mathcal{J}^{I}{}_{M} \tilde{\mathcal{J}}^{J}{}_{N}$ does not vanish, then
the Killing spinor must satisfy one condition Eq.~(\ref{eq:hyperprojector})
(with the upper sign only) for each value of $m$, each of which breaks the
supersymmetry a further factor of $1/2$ up to a maximum $1/(4N)$, which is the
fraction of supersymmetry preserved by a generic configuration.


\section{Equations of motion}
\label{sec-timelikeeom}

The supersymmetric configurations found in the previous section do not
necessarily satisfy all the equations of motion. In order to find
supersymmetric solutions, we have seen in Section~\ref{sec-kses} that it is
enough to require that the supersymmetric configurations satisfy the $0^{th}$
components of the Maxwell equations and Bianchi identities because the rest of
the equations of motion are then, according to the KSIs, automatically
satisfied. In this section we are going to find the 0th component of the
Maxwell equations and Bianchi identities and we will check that the KSIs are
satisfied for the supersymmetric configurations that we have obtained. This
will serve as a powerful cross-check of our results.

Let us start with the Maxwell equations and Bianchi identities: it is
convenient to construct a symplectic vector of 2-forms $\mathcal{F}$
containing the field strengths $F^{\Lambda}$ and their symplectic duals
$\tilde{F}_{\Lambda}$, by $\mathcal{F}^{T} \equiv \left( F^{\Lambda},
  \tilde{F}_{\Lambda}\right)$.  The Bianchi identities and Maxwell equations
can be written in the form $d\mathcal{F}=0$.

The field strengths $F^{\Lambda}$ can be easily deduced from the equations
obtained in Sec.~(\ref{sec-first}) and read

\begin{equation}
F^{\Lambda}  = F^{\Lambda +}+ F^{\Lambda -}
\equiv V^{-2}[\hat{V}\wedge E^{\Lambda} 
-\star (\hat{V}\wedge B^{\Lambda}) ]\, ,
\end{equation}

\noindent
where

\begin{equation}
  \begin{array}{rcl}
E^{\Lambda} & = & C^{\Lambda +} + C^{\Lambda +} = 
d (|M|^{2}\mathcal{R}^{\Lambda})\, ,\\
& & \\
B^{\Lambda} & = & -i(C^{\Lambda +} - C^{\Lambda +}) = 
-{\textstyle\frac{i}{2}}\left\{M^{IJ}\mathfrak{D}f^{\Lambda}{}_{IJ}
+f^{*\, \Lambda}{}_{IJ}\mathfrak{D}M^{IJ} -\mathrm{c.c.} \right\}\, ,\\
\end{array}
\end{equation}

Using the same results one can deduce

\begin{equation}
\tilde{F}_{\Lambda}  = \mathcal{N}^{*}_{\Lambda\Sigma}F^{\Sigma +}+ 
 \mathcal{N}_{\Lambda\Sigma}F^{\Lambda -}
\equiv V^{-2}[\hat{V}\wedge E_{\Lambda} 
-\star (\hat{V}\wedge B_{\Lambda}) ]\, ,
\end{equation}

\noindent
where

\begin{equation}
  \begin{array}{rcl}
E_{\Lambda} & = &  \mathcal{N}^{*}_{\Lambda\Sigma}C^{\Sigma +}+ 
 \mathcal{N}_{\Lambda\Sigma}C^{\Lambda -} = 
d (|M|^{2}\mathcal{R}_{\Lambda})\, ,\\
& & \\
B_{\Lambda} & = & -i(\mathcal{N}^{*}_{\Lambda\Sigma}C^{\Sigma +}- 
 \mathcal{N}_{\Lambda\Sigma}C^{\Lambda -}) =
-{\textstyle\frac{i}{2}}\left\{M^{IJ}\mathfrak{D}h_{\Lambda\, IJ}
+h^{*}{}_{\Lambda\, IJ}\mathfrak{D}M^{IJ} -\mathrm{c.c.} \right\}\, .\\
\end{array}
\end{equation}

\noindent
Combining the two expressions one can see that the symplectic vector $F$ is
given by

\begin{equation}
\label{eq:F}
\mathcal{F}\ =\ V^{-2}\left\{\hat{V} \wedge d
  (|M|^{2}\mathcal{R})   
+{\textstyle\frac{i}{2}}\left[\hat{V}\wedge 
\left(M^{IJ}\mathfrak{D}\mathcal{V}_{IJ}
+\mathcal{V}^{*\,  IJ}\mathfrak{D}M_{IJ} -\mathrm{c.c.}\right) \right]
\right\}\, .  
\end{equation}

\noindent
Using the equation for $\omega$ (\ref{eq:do}) and $\mathfrak{D}\mathcal{J}=0$,
it can be rewritten in the form

\begin{equation}
\mathcal{F}\ =\  -{\textstyle\frac{1}{2}} d (\mathcal{R}\hat{V})   
-{\textstyle\frac{1}{2}}\star
(\hat{V}\wedge 
d\mathcal{I}) 
\, ,  
\end{equation}

The combined Maxwell equations and Bianchi identities (\textit{i.e.}
$d\mathcal{F}=0$) then imply the equations

\begin{equation}
\label{eq:M+B}
d\star
(\hat{V}\wedge 
d\mathcal{I}) 
=0\, ,  
\end{equation}





\noindent
which, can be rewritten in the form

\begin{equation}
\label{eq:M+B2}
\mathcal{E}^{a}=
{\textstyle\frac{1}{\sqrt{2}}|M|}
\delta^{a}{}_{0}\nabla^{2}_{(3)}\mathcal{I} 
=0\, ,
\end{equation}

\noindent
in full agreement with the fact, derived from the KSIs, that the Maxwell and
Bianchi equations only have nontrivial $0^{th}$ component.

To calculate $\mathcal{E}_{00}$ we need to use Eq.~(\ref{eq:MRI}) to express
the second derivatives of $|M|$ in terms of symplectic sections. Then

\begin{equation}
-\nabla^{2}\langle\,\mathcal{R}\mid \mathcal{I}\, \rangle=
2\langle\,\nabla^{2}\mathcal{I}\mid \mathcal{R}\, \rangle
+2\langle\,\nabla_{m}\mathcal{I}\mid \nabla_{m}\mathcal{R}\, \rangle\, .  
\end{equation}

\noindent
Using in the second term Eq.~(\ref{eq:partialcontractionvielbeine1}) we find
that

\begin{equation}
\label{eq:E00}
  \begin{array}{rcl}
\mathcal{E}_{00} & = & G_{00} 
+{\textstyle\frac{1}{24}}\alpha_{1}P^{*\, IJKL}{}_{m}P_{IJKL}{}_{m}
+{\textstyle\frac{1}{2}}\alpha_{2}P^{*\, iIJ}{}_{m}P_{iIJ\, m}
-8\Im {\rm m}\mathcal{N}_{\Lambda\Sigma}
F^{\Lambda\, +}{}_{0m}F^{\Sigma\, -}{}_{0m}\\
& & \\
& = & 
-2|M|^{4}\langle\,\nabla_{(3)}^{2}\mathcal{I}\mid \mathcal{R}\, \rangle
+{\textstyle\frac{1}{2}}|M|^{2}
\left[R(\gamma) +6|M|^{-2}\Pi^{IJ}{}_{KL}
\mathfrak{D}_{m}M_{IJ}\mathfrak{D}_{m}M^{KL}
\right.
\\
& & \\
& & 
+{\textstyle\frac{1}{12}}
\alpha_{1}\left(\delta^{IJ}{}_{KL}
-6\alpha_{1}^{-1}|M|^{-2}M^{IJ}M_{KL} \right) P_{IJMN\, m}P^{*\, KLMN}{}_{m}
\\
& & \\
& & 
\left.
+\alpha_{2}\left(\delta^{IJ}{}_{KL}-\alpha_{2}^{-1}|M|^{-2}M^{IJ}M_{KL}\right)
P_{iIJ\,  m}P^{*\, iKL}{}_{m}
\right]\, .
\end{array}
\end{equation}

It is straightforward to show that $\mathcal{E}_{0m}=0$ identically, and, for
simplicity, we compute

\begin{equation}
\label{eq:tracelessEmn}
  \begin{array}{rcl}
|M|^{-2}[\mathcal{E}_{mn} +\tfrac{1}{2}\delta_{mn}\mathcal{E}_{\mu}{}^{\mu}] 
& = & 
-\frac{\sqrt{2}}{|M|^{3}}\langle\,\mathcal{E}^{0} \mid \mathcal{R}\, \rangle
+R(\gamma)_{mn}
-2|M|^{-2}\Pi^{IJ}{}_{KL}\mathfrak{D}_{(m|}M_{IJ}\mathfrak{D}_{|n)}M^{KL}
\\
& & \\
& & 
+{\textstyle\frac{1}{12}}
\alpha_{1}\left(\delta^{IJ}{}_{KL}
-6\alpha_{1}^{-1}|M|^{-2}M^{IJ}M_{KL} \right) 
P_{IJMN\ (m}P^{*\ KLMN}{}_{n)}
\\
& & \\
& & 
+\alpha_{2}\left(\delta^{IJ}{}_{KL}-\alpha_{2}^{-1}|M|^{-2}M^{IJ}M_{KL}\right)
P_{iIJ\ (m}P^{*\ iKL}{}_{n)}
\, .
\end{array}
\end{equation}

Finally, from Eqs.~(\ref{eq:cij+}) and (\ref{eq:ti+})  we find that the
scalar equations of motion are given by:

\begin{description}
\item[$N=2$::]

  \begin{equation}
\label{eq:scalareomN=2}
    \begin{array}{rcl}
      -|M|^{-2}\mathcal{E}^{iIJ} & = & 
      \mathfrak{D}_{m} P^{*\ iIJ}{}_{m} -2|M|^{-2}\mathfrak{D}_{m}M^{IJ} 
      M_{KL} P^{*\, i KL}{}_{m}
      \\
      & & \\
      & & 
      -\tfrac{1}{2}|M|^{-2}P^{*\, iIJ\, A}P^{*\, jk}{}_{A}
      M^{KL}M^{MN} P_{jKL\, m}P_{kMN\, m}
      \, .\\  
    \end{array}
  \end{equation}

\item[$N=3$::]

\begin{equation}
\label{eq:scalareomN=3}
 -|M|^{-2}\mathcal{E}^{iIJ} =
\mathfrak{D}_{m} P^{*\ iIJ}{}_{m} -2|M|^{-2}
\mathfrak{D}_{m}M^{IJ}   M_{KL} P^{*\, i KL}{}_{m}\, ,
\end{equation}

\noindent
or, in terms of the dual variables

\begin{equation}
-|M|^{-2}\tilde{\mathcal{E}}^{i}{}_{I} =
\mathfrak{D}_{m} \tilde{P}^{i}{}_{I\, m} -2|\tilde{M}|^{-2}
\mathfrak{D}_{m}\tilde{M}_{I}   \tilde{M}^{J} \tilde{P}^{i}{}_{J\, m}\, .
\end{equation}

\item[$N=4$::]

\begin{equation}
  \begin{array}{rcl}
    -|M|^{-2}\mathcal{E}^{IJKL} & = & 
    \mathfrak{D}_{m} P^{*\ IJKL}{}_{m}
    -12|M|^{-2}
      M_{MN}P^{*\, MN[IJ|}{}_{m}\mathfrak{D}_{m}M^{|KL]} 
    \\
    & &  \\
    & & 
    -\tfrac{1}{2}|M|^{-2}P^{*\, IJKL\, A}P^{*\, ij}{}_{A}
    M^{MN}M^{PQ} P_{i\, MN\, m}P_{i\, PQ\, m}
    \, ,\\
  \end{array}
\end{equation}

\noindent
or

\begin{equation}
\label{eq:scalareomN=4-1}
    -|M|^{-2}\mathcal{E} =
    \mathfrak{D}_{m} P_{m}
    -2|M|^{-2} M_{IJ}\mathfrak{D}_{m}M^{IJ}P_{m} 
    -\tfrac{1}{2}|M|^{-2} M^{MN}M^{PQ} P_{iMN\, m}P_{jPQ\, m}
    \, ,
\end{equation}

\noindent
and

\begin{equation}
\label{eq:scalareomN=4-2}
  \begin{array}{rcl}
    -|M|^{-2}\mathcal{E}^{iIJ} & = & 
    \mathfrak{D}_{m} P^{*\ iIJ}{}_{m} -2|M|^{-2}\left[
      \mathfrak{D}_{m}M^{IJ} 
      +{\textstyle\frac{1}{2}}
      M_{MN}P^{*\, MNIJ}{}_{m} \right]
    M_{KL} P^{*\, i KL}{}_{m}
    \\
    & & \\
    & & 
    -|M|^{-2}\varepsilon^{IJKL}\left[
      \mathfrak{D}_{m}M_{KL} 
      +{\textstyle\frac{1}{2}}
      M^{MN}P_{MNKL\, m} \right]
    M^{PQ} P_{iPQ\, m}
    \, .\\  
  \end{array}
\end{equation}

\item[$N=5$::]

\begin{equation}
 -|M|^{-2}\mathcal{E}^{IJKL} =
 \mathfrak{D}_{m} P^{*\ IJKL}{}_{m}
 -12|M|^{-2}M_{MN}P^{*\, MN[IJ|}{}_{m}\mathfrak{D}_{m}M^{|KL]}   \, ,
\end{equation}

\noindent 
or

\begin{equation}
\label{eq:scalareomN=5}
 -|M|^{-2}\tilde{\mathcal{E}}_{I} =
 \mathfrak{D}_{m} \tilde{P}_{I\, m}
 -2|M|^{-2}\mathfrak{D}_{m}\tilde{M}_{IJK}\tilde{M}^{JKL}P_{L\, m}\, .
\end{equation}

\item[$N=6$::]

\begin{equation}
  \begin{array}{rcl}
    -|M|^{-2}\mathcal{E}^{IJKL} & = & 
    \mathfrak{D}_{m} P^{*\ IJKL}{}_{m}
    -12|M|^{-2}\left[
      M_{MN}P^{*\, MN[IJ|}{}_{m}\mathfrak{D}_{m}M^{|KL]} 
    \right. \\
    & &  \\
    & & 
    \left.
      +{\textstyle\frac{1}{4}} M_{MN}P^{*\, MN[IJ|}{}_{m}
      M_{OP}P^{*\, OP|KL]}{}_{m} \right]
    \\
    & &  \\
    & & 
    -|M|^{-2}\varepsilon^{IJKLMN}
    \left[\mathfrak{D}_{m}M_{MN} 
      +{\textstyle\frac{1}{2}}
      M^{PQ}P_{PQMN\, m} \right]
    M^{RS} P_{RS\, m}
    \, ,\\
  \end{array}
\end{equation}

\noindent
or

\begin{equation}
\label{eq:scalareomN=6}
  \begin{array}{rcl}
    -|M|^{-2}\mathcal{E}_{IJ} & = & 
    \mathfrak{D}_{m} P_{IJ\, m}
    -|M|^{-2}\left[ \mathfrak{D}_{m}\tilde{M}_{IJKL} 
      +{\textstyle\frac{1}{2}}M_{IJ}P_{KL\, m} \right] \tilde{M}^{KLMN}
P_{MN\,  m} 
    \\
    & &  \\
    & & 
    -2|M|^{-2}
    \left[\mathfrak{D}_{m}M_{IJ} 
      +{\textstyle\frac{1}{2}}\tilde{M}_{IJKL}P^{*\, KL}{}_{m} \right]
    M^{RS} P_{RS\, m}
    \, ,\\
  \end{array}
\end{equation}

\noindent 
and finally

\item[$N=8$::]

\begin{equation}
\label{eq:scalareomN=8}
  \begin{array}{rcl}
    -|M|^{-2}\mathcal{E}^{IJKL} & = & 
    \mathfrak{D}_{m} P^{*\ IJKL}{}_{m}   \\
    & &  \\
    & &  
\hspace{-3cm}
-12|M|^{-2}\left[
            M_{MN}P^{*\, MN[IJ|}{}_{m}\mathfrak{D}_{m}M^{|KL]} 
      +{\textstyle\frac{1}{4}} M_{MN}P^{*\, MN[IJ|}{}_{m}
      M_{OP}P^{*\, OP|KL]}{}_{m} \right]
    \\
    & &  \\
    & & 
\hspace{-3cm}
    -\tfrac{1}{2}|M|^{-2}\varepsilon^{IJKLMNPQ}
    \left[
      M^{RS}P_{RS[MN|\, m}\mathfrak{D}_{m}M_{|PQ]} 
      +{\textstyle\frac{1}{4}} M^{RS}P_{RS[MN|\, m}
      M^{TU}P_{TU|PQ]\, m} \right]
    \, .\\
  \end{array}
\end{equation}

\end{description}

\subsection{Checking the KSIs}
\label{sec-timelikeksischeck}

Let us start by checking the KSI Eq.~(\ref{eq:timelikeksi2-2}). Substituting
the above expression, we get

\begin{equation}
\langle\, \nabla^{2}_{(3)}\mathcal{I} 
\mid\, \mathcal{I} \rangle
=0\, .
\end{equation}

\noindent 
The r.h.s.~vanishes identically due to the integrability condition of
the equation that defines the 1-form $\omega$, Eq.~(\ref{eq:oidiintcond}),
whose existence is a necessary condition of supersymmetry. 

To check the KSI Eq.~(\ref{eq:timelikeksi2-1}) we need to compute
$\langle\,\mathcal{E}^{0} \mid \mathcal{R}\, \rangle$:

\begin{equation}
\langle\,\mathcal{E}^{0} \mid \mathcal{R}\, \rangle 
=
{\textstyle\frac{1}{\sqrt{2}}} |M|^{3}
\langle\, \nabla_{(3)}^{2}\mathcal{I} \mid \mathcal{R}\, \rangle\, .
\end{equation}

\noindent
Comparing this with the expression for $\mathcal{E}_{00}$ given in
Eq.~(\ref{eq:E00}) we find that supersymmetry, requires the following relation
between the curvature of the 3-dimensional space and the scalars

\begin{equation}
\label{eq:susyconditiononR}
  \begin{array}{rcl}
R(\gamma) & = & 
-{\textstyle\frac{1}{12}}
\alpha_{1}\left(\delta^{IJ}{}_{KL}
-6\alpha_{1}^{-1}\mathcal{J}^{I}{}_{K}\mathcal{J}^{I}{}_{L} \right) P_{IJMN\, m}P^{*\, KLMN}{}_{m}
\\
& & \\
& & 
-\alpha_{2}\left(\delta^{IJ}{}_{KL}-\alpha_{2}^{-1}
\mathcal{J}^{I}{}_{K}\mathcal{J}^{I}{}_{L}\right)
P_{iIJ\,  m}P^{*\, iKL}{}_{m}
\, ,
\end{array}
\end{equation}

\noindent
a result we will comment upon shortly.

As for the KSI (\ref{eq:timelikeksi1}) we point out that, as we mentioned in the previous
section, $\mathcal{E}_{0m}$ vanishes identically; from
Eq.~(\ref{eq:tracelessEmn}) we see that $\mathcal{E}_{mn}$ vanishes if
Eq.~(\ref{eq:susyconditiononR}) is satisfied and furthermore that

\begin{equation}
\label{eq:susyconditiononRxy}
  \begin{array}{rcl}
R(\gamma)_{mn} & = & 
-{\textstyle\frac{1}{12}}
\alpha_{1}\left(\delta^{IJ}{}_{KL}
-6\alpha_{1}^{-1}\mathcal{J}^{I}{}_{K}\mathcal{J}^{I}{}_{L}\right) 
P_{IJMN\ (m}P^{*\ KLMN}{}_{n)}
\\
& & \\
& & 
-\alpha_{2}\left(\delta^{IJ}{}_{KL}
-\alpha_{2}^{-1}\mathcal{J}^{I}{}_{K}\mathcal{J}^{I}{}_{L}\right)
P_{iIJ\ (m}P^{*\ iKL}{}_{n)}
\, .
\end{array}
\end{equation}

\noindent
This is the only equation we really need to impose on the 3-dimensional metric
as Eq.~(\ref{eq:susyconditiononR}) is nothing but its trace. One can show
(case by case, for each $N$) that this expression is completely equivalent to
Eqs.~(\ref{eq:Rg}), which are satisfied by the supersymmetric configurations.

We can then check those KSIs that relate the equations of motion of the
scalars to the $0^{th}$ component of the Maxwell and Bianchi equations. It is
convenient to first compute them for the result for a generic value of $N$,
and then consider a specific value.  {}For generic $N$ one obtains

\begin{equation}
  \begin{array}{rcl}
\langle\,\mathcal{E}^{0} \mid \mathcal{V}^{*\, i}\, \rangle &  = &   
{\textstyle\frac{1}{2\sqrt{2}}}|M|
\left\{\mathfrak{D}_{m} P^{*\ iIJ}{}_{m} M_{IJ}
-2 |M|^{-2}P^{*\ iIJ}{}_{m} M_{IJ} M_{KL}
\mathfrak{D}_{m}M^{KL}
\right. \\
& & \\
& & 
- M^{IJ}\left[P_{jIJ\, m}P^{*\, ij}{}_{m} 
+\tfrac{1}{2}P_{IJKL\, m}P^{*\, iKL}{}_{m}\right] 
\biggr\}\, . 
\end{array}
\end{equation}

\noindent
and 

\begin{equation}
  \begin{array}{rcl}
\langle\,\mathcal{E}^{0} \mid \mathcal{V}^{*\, IJ}\, \rangle &  = &   
{\textstyle\frac{1}{2\sqrt{2}}}|M|
\biggl\{\mathfrak{D}_{m} P^{*\ IJKL}{}_{m} M_{KL}
-2|M|^{-2} \mathfrak{D}_{m} P^{*\ IJKL}{}_{m} M_{KL}M_{MN}\mathfrak{D}_{m}M^{MN}
\\
& & \\
& & 
-\tfrac{1}{2}M^{MN}\left[P^{*\, IJKL}{}_{m} P_{KLMN\, m}
+2P^{*\, i\, IJ}{}_{m} P_{i\, MN\, m}\right] 
\biggr\}\, .
\end{array}
\end{equation}

\noindent
\begin{itemize}
\item[$\mathbf{N=2}$::] it is enough to check the KSI Eq.~(\ref{eq:ksi3-1})
  using the form of the equation of motion derived before
  Eq.~(\ref{eq:scalareomN=2}) being careful with the $P^{2}$ and $P^{4}$
  terms. A detailed calculation shows that they cancel each other, in
  agreement with the results of Ref.~\cite{Meessen:2006tu}.

\item[$\mathbf{N=3}$::] For the case $N=3$ we have to check the KSI
  Eq.~(\ref{eq:scalarKSIN=3}) using the form of the equation of motion derived
  before Eq.~(\ref{eq:scalareomN=3}). Again, it is readily found to be
  satisfied by using the condition Eq.~(\ref{eq:mattersusycondition4}) and the
  covariant constancy of $\mathcal{J}$.

\item[$\mathbf{N=4}$::] For the case $N=4$ we have to check the KSIs
  Eqs.~(\ref{eq:scalarKSIN=4-1}) and (\ref{eq:scalarKSIN=4-2}) using
  Eqs.~(\ref{eq:scalareomN=4-1}) and Eq.~(\ref{eq:scalareomN=4-2})
  respectively. The first KSI is easily seen to be satisfied. The second KSI
  is satisfied up to a term of the form

\begin{equation}
\label{eq:eomhypers}
\mathfrak{D}_{m}(P_{i\, MN\, m}
\mathcal{J}^{M}{}_{[I}\tilde{\mathcal{J}}^{N}{}_{J]})\, ,  
\end{equation}

\noindent
which vanishes automatically after use of the constraint
Eqs.~(\ref{eq:PiIJcondition}) and (\ref{eq:ds=0}). This term can be seen as
the equation of motion for the hypers of the associated $N=2$ truncation and,
as it happens in the $N=2$ theory, it is automatically satisfied for the
supersymmetric configurations independently of whether the Maxwell equations
and Bianchi identities are satisfied or not.

\item[$\mathbf{N=5}$::] For the case $N=5$ we have to check the KSI
  Eq.~(\ref{eq:scalarKSIN=5}) using Eq.~(\ref{eq:scalareomN=5}). In this case
  the crucial property that makes it to be satisfied is
  Eq.~(\ref{eq:mattersusycondition3}).

\item[$\mathbf{N=6}$::] In the $N=6$ case we find the the KSI
  Eq.~(\ref{eq:scalarKSIN=6}) is satisfied Eq.~(\ref{eq:scalareomN=6}) up to a
  term of the form Eq.~(\ref{eq:eomhypers}), which is also seen to vanish
  identically.

\item[$\mathbf{N=8}$::] Finally, in the $N=8$ case we find the the KSI
  Eq.~(\ref{eq:scalarKSIN=8}) is satisfied Eq.~(\ref{eq:scalareomN=8}) up to a
  term of the form

\begin{equation}
\label{eq:eomhypers2}
\mathfrak{D}_{m}(P_{IJKL\, m}\, 
\mathcal{J}^{I}{}_{[M}
\tilde{\mathcal{J}}^{J}{}_{N}
\tilde{\mathcal{J}}^{K}{}_{P}
\tilde{\mathcal{J}}^{L}{}_{Q]})\, ,  
\end{equation}

\noindent
which vanishes upon use of Eqs.~(\ref{eq:PIJKLcondition}) and (\ref{eq:ds=0}).

\end{itemize}
\noindent
In conclusion we see that the KSIs are always satisfied.


\section{Conclusions}
\label{sec-conclusions}

The results presented in this paper are a first step towards a full
characterization of all the four-dimensional supersymmetric solutions preserving at least
one supercharge. It is
clear that further work is needed in order to make the general solutions
presented here more explicit for each $N$: first of all, convenient
parametrizations of the matrices $M^{IJ}$ satisfying all the required
properties (in particular all the supersymmetry constraints involving the
projector $\mathcal{J}$) and general ways to construct the generalized Pauli
matrices $\sigma^{m}$ have to be found, the stabilization equations have to be
solved (this is in general hard, and might prove impossible); furthermore, the
scalar fields need to be resolved; the would-be vector-scalars should be
resolved in terms of the harmonic functions and the would-be hyperscalars
should be found the hard way by solving the relevant equations
(\ref{eq:PIJKLcondition},\ref{eq:PiIJcondition}) and their consistent
interplay with the connection on the 3-dimensional base space,
Eq.~(\ref{eq:3dconnection}). Only then will we have explicit expressions for
the supersymmetric solutions. The problem is similar to, but definitely more
involved than, finding supersymmetric solutions in $d=4$ $N=2$ supergravities
coupled to vector and hypermultiplets \cite{Huebscher:2006mr}. A further issue
that needs to be investigated and which does not arise in the $N=2$ $d=4$ case
is the classification of supersymmetric solutions preserving more than the
minimal amount of supersymmetry.

The supersymmetric black hole solutions of the 4-dimensional supergravities
are a very interesting subclass of the supersymmetric solutions identified
here.  They are ``hyper-less'' (\textit{i.e.} they have a flat 3-dimensional
base space) solutions and, therefore, simpler to construct. The black-hole
solutions of $N=8$ are particularly interesting due to the possible
ultraviolet-finiteness of the theory, \textit{e.g.} \cite{Bern:2009kd}. There
are many partial results in the literature
\cite{Khuri:1995xk,Arcioni:1998mn,Bertolini:1998mt,Bertolini:1999je} including
very large families of solutions obtained via $N=2$ truncations of the theory
\cite{Ferrara:2006yb} but the derivation of a manifestly $E_{7(7)}$-invariant
family of solutions on which the conjectures concerning the
$E_{7(7)}$-invariant entropy formula \cite{Kallosh:1996uy} could be explicitly
checked is highly desirable. Our results provide a starting point for this
derivation \cite{kn:MOS}.

The attractor mechanism \cite{Ferrara:1995ih} (see also the more recent
reference \cite{Ceresole:2009jc}) has been one of the main tools for the study
of supersymmetric black-hole solutions. Our results establish a clear
distinction between the scalars which are driven by the electric and magnetic
charges of the vector fields (which would belong to the would-be vector
multiplets of the associated $N=2$ truncation) and, therefore, subject to the
attractor mechanism, and those that are not (which would belong to the
would-be hypermultiplets of the associated $N=2$ truncation). A simple
derivation of the attractor flow equations for the first kind of scalars based
on the general form of the solutions found here can be readily given
\cite{kn:O}.

Another interesting class of timelike supersymmetric solutions which deserves
to be studied in more detail is the class of domain walls associated to the
supersymmetry projectors $\Pi^{m\pm\, I}{}_{J}$ and, therefore, to the
would-be hyperscalars of the associated $N=2$ truncation.

Finally, to complete the program of characterizing all supersymmetric solutions, the supersymmetric solutions in the null
class need to be identified. In the null class the $\mathrm{U}(N)$
R-symmetry group is broken to $\mathrm{U}(1)\times \mathrm{U}(N-1)$ and there
is an ''$N=1$ truncation'' associated to the $\mathrm{U}(1)$ subgroup
\cite{kn:HMOV}. The solutions will then be analogous to the supersymmetric
solutions of the ungauged $N=1$ theories with no superpotential, classified in
Refs.~\cite{art:16th} and \cite{Ortin:2008wj}, and include waves, strings and domain walls.


\section*{Acknowledgments}

This has been supported in part by the Spanish Ministry of Science and
Education grants FPA2006-00783 and FPA2009-07692, the Comunidad de Madrid
grant HEPHACOS S2009ESP-1473, the Spanish Consolider-Ingenio 2010 program CPAN
CSD2007-00042, a Ram\'on y Cajal fellowship RYC-2009-05014 (PM), the
Princip\'au d'Asturies grant IB09-069 (PM) and a MEC Juan de la Cierva
scholarship (SV).  TO wishes to express his gratitude to the CERN Theory
Division for its hospitality during the crucial stages of this work and
M.M.~Fern\'andez for her permanent support.

\appendix

\section{Generic scalar manifolds}
\label{sec-scalars}
All the scalar manifolds can be described by a $\mathrm{Usp}(\bar{n},\bar{n})$
matrix $U$ which is constructed in terms of the matrices\footnote{When we
  multiply these matrices we must include a factor $1/2$ for each contraction
  of pairs of antisymmetric indices $IJ$.}

\begin{equation}
f\equiv (f^{\Lambda}{}_{IJ}, f^{\Lambda}{}_{i})\, ,
\hspace{1cm}  
h\equiv (h_{\Lambda\, IJ}, h_{\Lambda\, i})\, ,
\end{equation}

\noindent 
where $I,J=1,\ldots N$ are the graviton-supermultiplet, or equivalently
$\mathrm{U}(N)$, indices and $i(=1,\ldots n)$ are indices labeling the vector
multiplets, and the embedding then imposes that $\bar{n}=n+N(N-1)/2$; this
information is represented in the following table:\footnote{ Observe that
  $N=6$ has $n=1$, even though there are no vector supermultiplets in this
  case.  This will be explained in Appendix (\ref{sec-multiplets}).  }
\begin{table}[!h]
\begin{center}
\begin{tabular}{|c||c|c|c|c|c|}
\hline
$N$&3&4&5&6&8\\\hline\hline
$n$&$n$&$n$&0&1&0\\\hline
$\bar{n}$ & $n+3$ &$n+6$&10&16&28\\\hline
\end{tabular}
\end{center}
\label{tablealphabeta}
\end{table} 

Using the above matrices one can then embed the generic scalar manifolds as

\begin{equation}
\label{eq:U}
U \equiv 
{\textstyle\frac{1}{\sqrt{2}}}
\left(
  \begin{array}{cc}
f+ih & f^{*}+ih^{*} \\
f-ih & f^{*}-ih^{*} \\
  \end{array}
\right)\, .  
\end{equation}

The condition that  $U\in \mathrm{Usp}(\bar{n},\bar{n})$ 

\begin{equation}
  \begin{array}{rcl}
U^{-1} & = & \left( 
\begin{array}{cc}
1 & 0 \\ 0 & -1 \\ 
\end{array}
\right)
U^{\dagger}
\left( 
\begin{array}{cc}
1 & 0 \\ 0 & -1 \\ 
\end{array}
\right)
 = 
\left( 
\begin{array}{cc}
0 & 1 \\ -1 & 0 \\ 
\end{array}
\right)
U^{T}
\left( 
\begin{array}{cc}
0 & -1 \\ 1 & 0 \\ 
\end{array}
\right)
\\
& & \\
& = & 
{\textstyle\frac{1}{\sqrt{2}}}
\left(
  \begin{array}{cc}
f^{\dagger}-ih^{\dagger} &  -(f^{\dagger}+ih^{\dagger}) \\
-(f-ih ) &  f+ih \\
  \end{array}
\right)\, , \\
\end{array}
\end{equation}

\noindent
leads to the following conditions for $f$ and $h$:

\begin{equation}
\label{eq:fhnormalization}
i(f^{\dagger}h-h^{\dagger}f) = 1\, ,
\hspace{1cm}
f^{T}h-h^{T}f=0\, .
\end{equation}

In terms of the symplectic vectors 

\begin{equation}\label{eq:symsec}
\mathcal{V}_{IJ}= 
\left(
  \begin{array}{c}
f^{\Lambda}{}_{IJ} \\ h_{\Lambda IJ} \\
  \end{array}
\right)\, ,  
\hspace{1cm}
\mathcal{V}_{i}= 
\left(
  \begin{array}{c}
f^{\Lambda}{}_{i} \\ h_{\Lambda\, i} \\
  \end{array}
\right)\, ,  
\end{equation}

\noindent
these constraints take the form\footnote{We use the convention 
\begin{equation}
\langle  \mathcal{A}\mid \mathcal{B}\rangle \equiv
  \mathcal{B}^{\Lambda}\mathcal{A}_{\Lambda} 
-\mathcal{B}_{\Lambda}\mathcal{A}^{\Lambda}\, .
\end{equation}
}

\begin{equation}
  \begin{array}{rcl}
\langle \mathcal{V}_{IJ}\mid\mathcal{V}^{*\, KL}\rangle 
& = &   
-2i\delta^{KL}{}_{IJ}\, , \\
& & \\
\langle \mathcal{V}_{i}\mid\mathcal{V}^{*\, j}\rangle 
&  = &   
-i\delta_{i}{}^{j}\, , \\
\end{array}
\end{equation}

\noindent
with the rest of the symplectic products vanishing.

The left-invariant Maurer-Cartan 1-form can be split into the Vielbeine $P$ and
the connection $\Omega$ as follows:

\begin{equation}
\Gamma \equiv U^{-1}dU = 
\left(
  \begin{array}{cc}
\Omega & P^{*} \\
P & \Omega^{*} \\
  \end{array}
\right)\, .  
\end{equation}

Thus, the different components of the connection are

\begin{equation}
\label{eq:O}
\Omega =
\left(
  \begin{array}{cc}
\Omega^{KL}{}_{IJ} & \Omega^{j}{}_{IJ} \\
\Omega^{KL}{}_{i} & \Omega^{j}{}_{i} \\
  \end{array}
\right) 
=
\left(
  \begin{array}{cc}
i \langle d \mathcal{V}_{IJ}\mid\mathcal{V}^{*\, KL}\rangle & 
i \langle d \mathcal{V}_{IJ}\mid\mathcal{V}^{*\, j}\rangle \\
i \langle d \mathcal{V}_{i}\mid\mathcal{V}^{*\, KL}\rangle & 
i \langle d \mathcal{V}_{i}\mid\mathcal{V}^{*\, j}\rangle \\
  \end{array}
\right)\, ,
\end{equation}

\noindent
and those of the Vielbeine are 

\begin{equation}
\label{eq:P}
P =
\left(
  \begin{array}{cc}
P_{KLIJ} & P_{jIJ} \\
P_{KLi} & P_{ij} \\
  \end{array}
\right) 
=
\left(
  \begin{array}{cc}
-i \langle d \mathcal{V}_{IJ}\mid\mathcal{V}_{KL}\rangle & 
-i \langle d \mathcal{V}_{IJ}\mid\mathcal{V}_{j}\rangle \\
-i \langle d \mathcal{V}_{i}\mid\mathcal{V}_{KL}\rangle & 
-i \langle d \mathcal{V}_{i}\mid\mathcal{V}_{j}\rangle \\
  \end{array}
\right)\, .
\end{equation}

The period matrix $\mathcal{N}_{\Lambda\Sigma}$ is defined by

\begin{equation}
\label{eq:periodmatrix}
\mathcal{N}=hf^{-1}=\mathcal{N}^{T}\, ,   
\end{equation}

\noindent
which implies properties which should be familiar from the $N=2$ case: for
instance

\begin{equation}
\label{eq:loweringindices}
\mathfrak{D}h_{\Lambda}=
\mathcal{N}^{*}_{\Lambda\Sigma}\mathfrak{D}f^{\Lambda}\, ,
\hspace{1cm}
h_{\Lambda}=\mathcal{N}_{\Lambda\Sigma}f^{\Sigma}\, ,
\end{equation}

and 

\begin{equation} 
\label{eq:completeness}
-{\textstyle\frac{1}{2}}(\Im{\rm m}\mathcal{N})^{-1|\Lambda\Sigma}=
{\textstyle\frac{1}{2}}f^{\Lambda}{}_{IJ}f^{*\Sigma IJ}
+f^{\Lambda}{}_{i}f^{*\Sigma\, i}\, ,
\end{equation}

\noindent
which can be derived from the definition of $\mathcal{N}$ and
Eq.~(\ref{eq:fhnormalization}).

We also quote the completeness relation

\begin{equation}
\label{eq:completeness2}
{\textstyle\frac{1}{2}}
 \mid \mathcal{V}_{IJ} \rangle\langle\mathcal{V}^{*\, IJ}  \mid\,\,
-{\textstyle\frac{1}{2}}
 \mid \mathcal{V}^{*\, IJ} \rangle\langle\mathcal{V}_{IJ}  \mid\,\,
+ \mid \mathcal{V}_{i} \rangle\langle\mathcal{V}^{*\, i}  \mid\,\,
- \mid \mathcal{V}^{*\, i} \rangle\langle\mathcal{V}_{i}  \mid\,\,
=i\, .
\end{equation}

Defining the $H_{Aut}\times H_{Matter}$ covariant derivative
according to

\begin{equation}
\label{eq:coder}
\mathfrak{D}\mathcal{V}=d\mathcal{V}-\mathcal{V}\Omega\, ,
\end{equation}  

\noindent
and using Eq.~(\ref{eq:loweringindices}) we obtain from (\ref{eq:O})

\begin{equation} 
\Omega^{KL}{}_{i}=\Omega^{j}{}_{IJ}=0\, ,
\end{equation}

\noindent
and from (\ref{eq:P})

\begin{eqnarray}
\label{eq:Paut}
P_{IJKL} & = & 
-2f^{\Lambda}{}_{IJ}\Im{\rm m}\mathcal{N}_{\Lambda\Sigma}\
\mathfrak{D}f^{\Sigma}{}_{KL}\, ,\\
& & \nonumber \\
\label{eq:Pmatter}
P_{iIJ} & = & -2f^{\Lambda}{}_{i}\Im{\rm m}\mathcal{N}_{\Lambda\Sigma}\
\mathfrak{D}f^{\Sigma}{}_{IJ}\, ,\\
& & \nonumber \\
\label{eq:Pij}
P_{ij} & = & -2f^{\Lambda}{}_{i}\Im{\rm m}\mathcal{N}_{\Lambda\Sigma}\
\mathfrak{D}f^{\Sigma}{}_{j}\, .
\end{eqnarray}

\noindent
The above equation can be inverted to give

\begin{eqnarray} 
\label{eq:dflij}
\mathfrak{D} f^{\Lambda}{}_{IJ} & = & 
f^{*\, \Lambda i} P_{iIJ}
+{\textstyle\frac{1}{2}}f^{*\Lambda KL}P_{IJKL}\, , \\
& & \nonumber \\
\label{eq:dfli}
\mathfrak{D} f^{\Lambda}{}_{i} & = & 
f^{*\, \Lambda j} P_{ij}
+{\textstyle\frac{1}{2}}f^{*\Lambda IJ}P_{iIJ}\, , 
\end{eqnarray}

\noindent
using Eq.~(\ref{eq:completeness}).

The definition of the covariant derivative leads to the identities

\begin{equation}
\langle\, \mathfrak{D}\mathcal{V}  \mid \mathcal{V}^{*}\, \rangle  
= 0\, ,
\hspace{1cm}
\langle\, \mathfrak{D}\mathcal{V}  \mid \mathcal{V}\, \rangle  
=\langle\, d\mathcal{V}  \mid \mathcal{V}\, \rangle  
=i P\, .
\end{equation}

The inverse Vielbeine $P^{*\, IJKL},P^{*\, iIJ},P^{*\, ij}$, satisfy (here $A$
labels the physical fields)

\begin{equation}
\label{eq:inversevielbeine}
P^{*\, IJKL\, A}P_{MNOP\, A} = 4!\delta^{IJKL}{}_{MNOP}\, , 
\hspace{.5cm}   
P^{*\, iIJ\, A}P_{jKL\, A} = 2\delta^{i}{}_{j}\delta^{IJ}{}_{KL}\, . 
\end{equation}

\noindent 
Their crossed products vanish but their products with $P_{ij\, A}$ do not.

We find 

\begin{eqnarray}
\label{eq:partialcontractionvielbeine1}
\langle\,  \mathfrak{D}_{A}\mathcal{V}_{IJ} \mid 
\mathfrak{D}_{B}\mathcal{V}^{*\ KL}\, \rangle 
& = & 
{\textstyle\frac{i}{2}}P_{IJMN}{}_{A}P^{*\ KLMN}{}_{B}  
+iP_{i IJ}{}_{A}P^{*\ iKL}{}_{B}\, ,  
\\
& & \nonumber \\
\label{eq:partialcontractionvielbeine3}
\langle\,  \mathfrak{D}_{A}\mathcal{V}_{IJ} \mid 
\mathfrak{D}_{B}\mathcal{V}^{*\ i}\, \rangle 
& = & 
{\textstyle\frac{i}{2}}P_{IJKL}{}_{A}P^{*\ iKL}{}_{B}  
+iP_{jIJ}{}_{A}P^{*\ ij}{}_{B}\, ,  
\\
& & \nonumber \\
\label{eq:partialcontractionvielbeine2}
\langle\,  \mathfrak{D}_{A}\mathcal{V}_{i} \mid 
\mathfrak{D}_{B}\mathcal{V}^{*\ j}\, \rangle 
& = & 
{\textstyle\frac{i}{2}}P_{iIJ}{}_{A}P^{*\ iIJ}{}_{B}  
+iP_{ik}{}_{A}P^{*\ jk}{}_{B}\, ,  
\end{eqnarray}

\noindent
while $\langle\,  \mathfrak{D}_{A}\mathcal{V}_{IJ} \mid 
\mathfrak{D}_{B}\mathcal{V}_{KL}\, \rangle 
=
\langle\,  \mathfrak{D}_{A}\mathcal{V}_{IJ} \mid 
\mathfrak{D}_{B}\mathcal{V}_{i}\, \rangle 
=
\langle\,  \mathfrak{D}_{A}\mathcal{V}_{i} \mid 
\mathfrak{D}_{B}\mathcal{V}_{j}\, \rangle =0
$.

Using the definition of the period matrix Eq.~(\ref{eq:periodmatrix}),
equation (\ref{eq:loweringindices}) and the first of
Eqs.~(\ref{eq:fhnormalization}) we get

\begin{equation}
d\mathcal{N} = 4i \Im{\rm m}\mathcal{N}\, \mathfrak{D}f f^{\dagger}\,
\Im{\rm m}\mathcal{N}\, .
\end{equation}

\noindent
This expression can be expanded in terms of the Vielbeine, using
Eqs.~(\ref{eq:dflij}) and (\ref{eq:dfli})

\begin{equation}
d\mathcal{N}_{\Lambda_{\Sigma}} = 
i \Im{\rm m}\mathcal{N}_{\Gamma(\Lambda} \Im{\rm m}\mathcal{N}_{\Sigma)\Omega}
\left[ P_{IJKL}f^{*\, \Gamma IJ}f^{*\, \Omega KL} 
+4 P_{iIJ}f^{*\, \Gamma i}f^{*\, \Omega IJ}
+4 P_{ij}f^{*\, \Gamma i}f^{*\, \Omega j}
\right]\, .
\end{equation}

\noindent
and, using Eqs.~(\ref{eq:inversevielbeine}) and taking into account that their
contraction with $P_{ij}$ does not necessarily vanish, implies

\begin{eqnarray}
\label{eq:PIJKLdN}
P^{*\, IJKL\, A}\frac{\partial}{\partial\phi^{A}}\mathcal{N}_{\Lambda\Sigma} 
& = & 
4!i \Im{\rm m}\mathcal{N}_{\Omega(\Lambda}   
\Im{\rm m}\mathcal{N}_{\Sigma) \Delta} 
f^{*\, \Omega [IJ|} f^{*\, \Delta |KL]}\, ,\\
& & \nonumber \\
\label{eq:PiIJdN}
P^{*\, iIJ\, A}\frac{\partial}{\partial\phi^{A}}\mathcal{N}_{\Lambda\Sigma} 
& = & 
8i \Im{\rm m}\mathcal{N}_{\Omega(\Lambda}   
\Im{\rm m}\mathcal{N}_{\Sigma) \Delta} 
f^{*\, \Omega i} f^{*\, \Delta IJ}\, .\\
& & \nonumber \\
\label{eq:P*IJKLdN}
P^{*\, IJKL\, A}\frac{\partial}{\partial\phi^{A}}\mathcal{N}^{*}_{\Lambda\Sigma} 
& = & 
-4i \Im{\rm m}\mathcal{N}_{\Omega(\Lambda}   
\Im{\rm m}\mathcal{N}_{\Sigma) \Delta} 
P^{*\, IJKL\, A}P^{*\, ij}{}_{A}
f^{\Omega}{}_{i} f^{\Delta}{}_{j}\, ,\\
& & \nonumber \\
\label{eq:P*iIJdN}
P^{*\, iIJ\, A}\frac{\partial}{\partial\phi^{A}}\mathcal{N}^{*}_{\Lambda\Sigma} 
& = & 
-4i \Im{\rm m}\mathcal{N}_{\Omega(\Lambda}   
\Im{\rm m}\mathcal{N}_{\Sigma) \Delta} 
P^{*\, iIJ\, A}P^{*\, jk}{}_{A}
f^{\Omega}{}_{i} f^{\Delta}{}_{j}\, .
\end{eqnarray}

Using the Maurer-Cartan equations $d\Gamma+\Gamma\wedge \Gamma =0$ and direct
calculations we find that the curvatures of $\Omega^{KL}{}_{IJ}$ and
$\Omega^{j}{}_{i}$ are

\begin{eqnarray}
R^{KL}{}_{IJ} & = & 
d \Omega^{KL}{}_{IJ} +{\textstyle\frac{1}{2}}\Omega^{KL}{}_{MN}\wedge
\Omega^{MN}{}_{IJ} 
 \nonumber \\
& & \nonumber \\
& = &  
-{\textstyle\frac{1}{2}}P^{* KLMN}\wedge P_{MNIJ}
- P^{*i KL}\wedge P_{iIJ}
\label{eq:RO=PP}
\\
& & \nonumber \\
& = & 
-i\langle\, \mathfrak{D}\mathcal{V}_{IJ}\mid  
\mathfrak{D}\mathcal{V}^{*\, KL}\,
\rangle\, ,\\
& & \nonumber \\
R^{j}{}_{i} & = & d \Omega^{j}{}_{i} +\Omega^{j}{}_{k}\wedge
\Omega^{k}{}_{i}
= 
-{\textstyle\frac{1}{2}}P^{*\, jIJ}\wedge P_{iIJ}
- P^{*ik}\wedge P_{ik}
\\
& & \nonumber \\
& = &  
-i\langle\, \mathfrak{D}\mathcal{V}_{i} \mid  
\mathfrak{D}\mathcal{V}^{*\, j} \,
\rangle\, .
\end{eqnarray}

\noindent
The vanishing of the curvature of $\Omega^{i}{}_{IJ}$ leads to 

\begin{equation}
\label{eq:esaidentidad}
{\textstyle\frac{1}{2}}P_{IJKL}\wedge P^{*\, iKL}
+ P_{jIJ}\wedge P^{*\, ij}
=
-i\langle\,  \mathfrak{D}\mathcal{V}_{IJ} \mid 
\mathfrak{D}\mathcal{V}^{*\, i}\, \rangle =0\, .
\end{equation}


\section{Generic $N\geq2$, $d=4$ multiplets}
\label{sec-multiplets}
In this section we will spill out the field content of the relevant graviton-
and vector supermultiplet\footnote{ The information in this appendix taken
  from Ref.~\cite{Andrianopoli:1996ve}, but adapted to the notations of
  Ref.~\cite{Bellorin:2005zc}.  } by giving said field content in a table
followed by the possible constraints that apply for for each individual case.

\begin{center}
\begin{tabular}{|c||c|c|c|c||c|c|c|c|}
\hline
\multicolumn{9}{|c|}{$N=3$} \\\hline\hline
&$e^{a}{}_{\mu}$ & $\psi_{I\mu}$  &$A^{IJ}{}_{\mu}$& 
$\chi_{IJK}$ & $A^{i}{}_{\mu}$ & $\lambda_{iI}$ &
$\lambda_{iIJK}$ & $P_{iIJ\mu}$  \\\hline
$\sharp$ & 1 & 3 & 3 & 1& $n$ & $3n$& $n$ & $(3+3)n$   \\\hline
\end{tabular}
\end{center}

\begin{table}[!h]
\begin{center}
\begin{tabular}{|c||c|c|c|c|c||c|c|c|c|}
\hline
\multicolumn{10}{|c|}{$N=4$} \\\hline\hline
&$e^{a}{}_{\mu}$ & $\psi_{I\mu}$  &$A^{IJ}{}_{\mu}$& $\chi_{IJK}$ & 
$P_{IJKL\mu}$& $A^{i}{}_{\mu}$ & $\lambda_{iI}$ &$\lambda_{iIJK}$ & 
$P_{iIJ\mu}$  \\\hline
$\sharp$ & 1 & 4 & 6 & 4& 1+1& $n$ & $4n$& $4n$ & $(6+6)n$   \\\hline
\end{tabular}
\end{center}
\end{table} 

In order to recover the $N=4$ field content we have to impose 

\begin{eqnarray} 
P_{i\, IJ}
& = & 
{\textstyle\frac{1}{2}}\varepsilon_{IJKL} P^{*\, i\, KL}\, ,
\\
& & \nonumber \\
\lambda_{i\, I}
& = & 
\tfrac{1}{3!}\varepsilon_{IJKL}\lambda^{i\, JKL}\, .
\end{eqnarray}

\begin{table}[!h]
\begin{center}
\begin{tabular}{|c||c|c|c|c|c|c|}
\hline
\multicolumn{7}{|c|}{$N=5$} \\\hline\hline
&$e^{a}{}_{\mu}$ & $\psi_{I\mu}$  &$A^{IJ}{}_{\mu}$& $\chi_{IJK}$ & 
$\chi^{IJKLM}$& $P_{IJKL\mu}$ \\\hline
$\sharp$ & 1 & 5 & 10 & 10& 1& $5+5$    \\\hline
\end{tabular}
\end{center}
\end{table}

\begin{table}[!h]
\begin{center}
\begin{tabular}{|c||c|c|c|c|c|c|c|c|c|c|}
\hline
\multicolumn{11}{|c|}{$N=6$} \\\hline\hline
&$e^{a}{}_{\mu}$ & $\psi_{I\mu}$  &$A^{IJ}{}_{\mu}$& $\chi_{IJK}$ & 
$\chi^{IJKLM}$& $P_{IJKL\mu}$ &$A$&$\lambda_{I}$&$\lambda_{IJK}$& 
$P_{IJ}$\\\hline
$\sharp$ & 1 & 6 & 15 & 20& 6& $15+15$ & 1&  6& 20& $15+15$\\\hline
\end{tabular}
\end{center}
\end{table} 

The situation for the $N=6$ case is a little bit more involved. In spite of
the fact that for $N=6$ there are no vector multiplets, the graviton multiplet
is obtained from the ``general case'' Eq.~(\ref{Ngrav}) coupling an extra
``vector multiplet''. This is because the decomposition of ${\rm SO}^*(12)$
with respect to ${\rm SU}(6)$ produces a singlet (this is the "practical
reason" why Eq.~(\ref{Ngrav}) is not enough).  The presence of the singlet
comes together with the fact that ${\rm SO}^*(12)/{\rm U}(6)$ has a Special
Geometry structure.

In order to recover the $N=6$ field content we have to impose

\begin{eqnarray} 
\lambda_{I} 
& = & 
{\textstyle\frac{1}{5!}}\varepsilon_{IJKLMN}\chi^{JKLMN}\, ,
\\
& & \nonumber \\
\chi_{IJK} 
& = & 
{\textstyle\frac1{3!}} \varepsilon_{IJKLMN}\lambda^{LMN}\, ,
\\
& & \nonumber \\
P_{IJKL}
& = & 
{\textstyle\frac{1}{2}}\varepsilon_{IJKLMN}P^{*\, MN}\, .
\end{eqnarray}

\begin{table}[!h]
\begin{center}
\begin{tabular}{|c||c|c|c|c|c|c|}
\hline
\multicolumn{7}{|c|}{$N=8$} \\\hline\hline
&$e^{a}{}_{\mu}$ & $\psi_{I\mu}$  &$A^{IJ}{}_{\mu}$& $\chi_{IJK}$ & 
$\chi^{IJKLM}$& $P_{IJKL\mu}$ \\\hline
$\sharp$ & 1 & 8 & 28 & 56& 56& $70+70$    \\\hline
\end{tabular}
\end{center}
\end{table} 

In order to recover the $N=8$ field content we have to impose 

\begin{eqnarray} 
P_{IJKL}
& = & 
{\textstyle\frac{1}{4!}}\varepsilon_{IJKLMNOP} P^{*\, MNOP}{}\, ,
\\
& & \nonumber \\
\chi_{IJK}
& = & 
{\textstyle\frac{1}{5!}}\varepsilon_{IJKLMNOP}\chi^{LMNOP}\, .
\end{eqnarray}


\section{Gamma matrices and spinors}

We work with a purely imaginary representation

\begin{equation}
\gamma^{a\, *}= -\gamma^{a}\, ,  
\end{equation}

\noindent 
and our convention for their anti-commutator is

\begin{equation}
\{\gamma^{a},\gamma^{b}\}= +2\eta^{ab}\, .  
\end{equation}

Thus, 

\begin{equation}
\gamma^{0}\gamma^{a}\gamma^{0}=
\gamma^{a\, \dagger}= (\gamma^{a})^{ -1}=\gamma_{a}\, .
\end{equation}

The chirality matrix is defined by

\begin{equation}
\gamma_{5}\equiv -i\gamma^{0}\gamma^{1}\gamma^{2}\gamma^{3}
={\textstyle\frac{i}{4!}} \epsilon_{abcd}
\gamma^{a}\gamma^{b}\gamma^{c}\gamma^{d}\, ,
\end{equation}

\noindent
and satisfies

\begin{equation}
\gamma_{5}{}^{\dagger}=-\gamma_{5}{}^{*}=\gamma_{5}\, ,
\hspace{1cm}
(\gamma_{5})^{2}=1\, .
\end{equation}

With this chirality matrix, we have the identity

\begin{equation}
\label{eq:dualgammaidentityind4}
\gamma^{a_{1}\cdots a_{n}} =\frac{(-1)^{\left[n/2\right]}i}{(4-n)!}
\epsilon^{a_{1}\cdots a_{n}b_{1}\cdots b_{4-n}} \gamma_{b_{1}\cdots b_{4-n}}
\gamma_{5}\, .
\end{equation}

Our convention for Dirac conjugation is

\begin{equation}
\bar{\psi}=i\psi^{\dagger}\gamma_{0}\, .  
\end{equation}

Using the identity Eq.~(\ref{eq:dualgammaidentityind4}) the general
$d=4$ Fierz identity for \textit{commuting} spinors takes the form

\begin{equation}
\label{eq:Fierzidentities}
  \begin{array}{rcl}
(\bar{\lambda} M\chi) (\bar{\psi} N \varphi ) & = &
{\textstyle\frac{1}{4}} (\bar{\lambda} M N \varphi) (\bar{\psi} \chi )
+{\textstyle\frac{1}{4}} (\bar{\lambda} M \gamma^{a}N \varphi) 
(\bar{\psi} \gamma_{a}\chi ) 
-{\textstyle\frac{1}{8}} (\bar{\lambda} M \gamma^{ab}N \varphi) 
(\bar{\psi} \gamma_{ab}\chi )
\\
& & \\
& & 
-{\textstyle\frac{1}{4}} (\bar{\lambda} M \gamma^{a}\gamma_{5}N \varphi) 
(\bar{\psi} \gamma_{a}\gamma_{5}\chi )
+{\textstyle\frac{1}{4}} (\bar{\lambda} M \gamma_{5}N \varphi) 
(\bar{\psi}\gamma_{5}\chi )\, .\\
\end{array}
\end{equation}

We use 4-component chiral spinors whose chirality is related to the
position of the $\mathrm{SU}(4)$-index:

\begin{equation}
\gamma_{5}\chi_{I}=+ \chi_{I}\, ,
\hspace{1cm}
\gamma_{5}\psi_{\mu\, I}=- \psi_{\mu\, I}\, ,
\hspace{1cm}
\gamma_{5}\epsilon_{I}=- \epsilon_{I}\, .
\end{equation}

\noindent
Both chirality and position of the $\mathrm{SU}(4)$-index are reversed under
complex conjugation, \textit{e.g.}

\begin{equation}
\gamma_{5}\chi_{I}^{*} \equiv\gamma_{5}\chi^{I}= -\chi^{I}\, ,
\hspace{1cm}
\gamma_{5}\psi_{\mu\, I}^{*}\equiv
\gamma_{5}\psi_{\mu}{}^{I}=+\psi_{\mu}{}^{I}\, ,
\hspace{1cm}
\gamma_{5}\epsilon_{I}^{*}\equiv
\gamma_{5}\epsilon^{I}= +\epsilon^{I}\, .
\end{equation}

We take this fact into account when Dirac-conjugating chiral spinors:

\begin{equation}
\bar{\chi}^{I}\equiv i(\chi_{I})^{\dagger}\gamma_{0}\, ,
\hspace{.5cm}
\bar{\chi}^{I}\gamma_{5}=-\bar{\chi}^{I}\, ,\,\,\,\, {\rm etc.}
\end{equation}


\section{Fierz identities for bilinears}
\label{sec-Fierz}

Here we are going to work with an arbitrary number $N$ of chiral spinors.
Whenever there are special results for particular values of $N$, we will
explicitly say so. We should bear in mind that the maximal number of
independent chiral spinors is 2 and $N(>2)$ spinors cannot be linearly
independent at a given point.  This trivial fact has important consequences.

Given $N$ chiral commuting spinors $\epsilon_{I}$ and their complex
conjugates $\epsilon^{I}$ we can constructed the following bilinears
that are not obviously related via
Eq.~(\ref{eq:dualgammaidentityind4}):

\begin{enumerate}
\item A complex matrix of scalars

\begin{equation}
M_{IJ}\equiv \bar{\epsilon}_{I}\epsilon_{J}\, ,  
\hspace{1cm}
M^{IJ}\equiv \bar{\epsilon}^{I}\epsilon^{J}=(M_{IJ})^{*}\, ,
\end{equation}

\noindent
which is antisymmetric $M_{IJ}=-M_{JI}$.

\item A complex matrix of vectors

\begin{equation}
V^{I}{}_{J\, a}\equiv i\bar{\epsilon}^{I}\gamma_{a}\epsilon_{J}\, ,  
\hspace{1cm}
V_{I}{}^{J}{}_{a}\equiv i\bar{\epsilon}_{I}\gamma_{a}\epsilon^{J}
=(V^{I}{}_{J\, a})^{*}\, ,
\end{equation}

\noindent
which is Hermitean:

\begin{equation}
(V^{I}{}_{J\, a})^{*}=V_{I}{}^{J}{}_{a} = V^{J}{}_{I\, a}
=(V^{I}{}_{J\, a})^{T}\, .  
\end{equation}

\item A complex matrix of 2-forms

\begin{equation}
\label{eq:p}
\Phi_{IJ\, ab}\equiv \bar{\epsilon}_{I}\gamma_{ab}\epsilon_{J}\, ,  
\hspace{1cm}
\Phi^{IJ}{}_{ab}\equiv 
\bar{\epsilon}^{I}\gamma_{ab}\epsilon^{J}=(\Phi_{IJ\, ab})^{*}\, ,
\end{equation}

\noindent
which is symmetric in the $\mathrm{SU}(N)$ indices $\Phi_{IJ\, ab}=\Phi_{JI\,
  ab}$ and furthermore is imaginary anti-selfdual, \textit{i.e.}

\begin{equation}
\label{eq:p1}
{}^{\star}\!\Phi_{IJ\, ab}=-i\Phi_{IJ\, ab}\,\,\, \Rightarrow\,\,\,
\Phi_{IJ\, ab} = \Phi_{IJ}{}^{+}{}_{ab}\, .
\end{equation}

\noindent
As we are going to see, this matrix of 2-forms can be expressed
entirely in terms of the scalar and vector bilinears.

\end{enumerate}

It is straightforward to derive identities for the products of these
bilinears using the Fierz identity Eq.~(\ref{eq:Fierzidentities}).
First, the products of scalars:

\begin{eqnarray}
M_{IJ}M_{KL} & = & {\textstyle\frac{1}{2}}M_{IL}M_{KJ} 
- {\textstyle\frac{1}{8}}\Phi_{IL}\cdot\Phi_{KJ}\, ,\label{eq:mm1}\\
& & \nonumber \\
M_{IJ}M^{KL} & = & -{\textstyle\frac{1}{2}} V^{L}{}_{I}\cdot V^{K}{}_{J}
\label{eq:mm2}\, .
\end{eqnarray}

\noindent
From Eq.~(\ref{eq:mm1}) immediately follows

\begin{equation}
\label{eq:mm3}
M_{I[J}M_{KL]}=0\, ,  
\end{equation}

\noindent
which is a Pl\"ucker identity and implies that $\mathrm{rank}(M_{IJ})\leq 2$.





We can define the $\mathrm{SU}(N)$-dual of $M_{IJ}$

\begin{equation}
\tilde{M}^{I_{1}\cdots I_{N-2}}\equiv 
{\textstyle\frac{1}{2}}\varepsilon^{I_{1}\cdots I_{N-2}KL}M_{KL}\, ,  
\hspace{1cm}
\varepsilon^{1\cdots N}=\varepsilon_{1\cdots N}=+1\, ,
\end{equation}

\noindent
in terms of which we can express Eq.~(\ref{eq:mm3}) as

\begin{equation}
\tilde{M}_{IJ_{1}\cdots J_{N-3}} M^{IK}=0\, .  
\end{equation}

\noindent
From Eq.~(\ref{eq:mm2}) and the antisymmetry of $M$ immediately
follows

\begin{equation}
\label{eq:vv1}
V^{I}{}_{L}\cdot V^{K}{}_{J}= -V^{I}{}_{J}\cdot V^{K}{}_{L}
=-V^{K}{}_{L}\cdot V^{I}{}_{J}\, ,
\end{equation}

\noindent
which implies that all the vector bilinears $V^{I}{}_{J\, a}$ are null:

\begin{equation}
\label{eq:vv2}
V^{I}{}_{J}\cdot V^{I}{}_{J}=0\hspace{3cm}(\mbox{no sum!})\; ,  
\end{equation}

\noindent
On the other hand, from Eqs.~(\ref{eq:vv1}) and (\ref{eq:mm2}) it follows that
the real, $\mathrm{SU}(N)$-invariant combination of vectors $V_{a}\equiv
V^{I}{}_{I\, a}$ is always non-spacelike:

\begin{equation}
\label{eq:vv3}
V^{2}=-V^{I}{}_{J}\cdot V^{J}{}_{I} = 2 M^{IJ}M_{IJ}\geq 0\, .
\end{equation}

The products of $M$ with the other bilinears\footnote{We omit the
  product $M_{IJ}\Phi_{KL\, ab}$ which will not be used.} give

\begin{eqnarray}
M_{IJ}V^{K}{}_{L\, a} & = & {\textstyle\frac{1}{2}}M_{IL}V^{K}{}_{J\, a} 
+{\textstyle\frac{1}{2}}\Phi_{IL\, ba} V^{K}{}_{J}{}^{b}\, ,
\label{eq:mv1}\\
& & \nonumber \\
M_{IJ}\Phi^{KL}{}_{ab} & = & 
V^{L}{}_{I\, [a|} V^{K}{}_{J\, |b]} 
-{\textstyle\frac{i}{2}} \epsilon_{ab}{}^{cd}
V^{L}{}_{I\, c} V^{K}{}_{J\, d} 
\label{eq:mp1}\, .
\end{eqnarray}

Now, let us consider the product of two arbitrary vectors\footnote{The product
  $V^{I}{}_{J\, a} V_{L}{}^{K}{}_{b}$ gives a different identity that will not
  be used}:

\begin{equation}
\label{eq:vv4}
V^{I}{}_{J\, a} V^{K}{}_{L\, b}
= {\textstyle\frac{i}{2}} \epsilon_{ab}{}^{cd}
V^{I}{}_{L\, c} V^{K}{}_{J\, d} 
+V^{I}{}_{L\, (a|} V^{K}{}_{J\, |b)}
-{\textstyle\frac{1}{2}} g_{ab}
V^{I}{}_{L}\cdot V^{K}{}_{J}\, .
\end{equation}

\noindent
For $V^{2}$ this identity allows us to write the metric in the form

\begin{equation}
\label{eq:vv5}
g_{ab}=2V^{-2}[V_{a}V_{b}-V^{I}{}_{J\, a} V^{J}{}_{I\, b}]\, .
\end{equation}

Following Tod \cite{Tod:1995jf}, for $V^{2}\neq 0$ we  introduce

\begin{equation}
\label{eq:J}
\mathcal{J}^{I}{}_{J} \equiv \frac{2 M^{IK} M_{JK}}{|M|^{2}}
= \frac{2 V\cdot V^{I}{}_{J}}{V^{2}}\, ,
\hspace{1cm}
|M|^{2}\equiv M^{LM}M_{LM}={\textstyle\frac{1}{2}}V^{2}\, .
\end{equation}

\noindent
Using Eq.~(\ref{eq:mm1}) we can show that it is a Hermitean projector
whose trace equals 2:

\begin{equation}
\label{eq:J1}
\mathcal{J}^{I}{}_{J}\mathcal{J}^{J}{}_{K}=
\mathcal{J}^{I}{}_{K}\, ,
\hspace{1cm}
\mathcal{J}^{I}{}_{I}=+2\, .
\end{equation}

\noindent
Further, using the general Fierz identity we find 

\begin{equation}
\label{eq:J2}
\mathcal{J}^{I}{}_{J}\epsilon^{J} =\epsilon^{I}\, ,  
\hspace{1cm}
\epsilon_{I}\mathcal{J}^{I}{}_{J} =\epsilon_{J}\, ,  
\end{equation}

\noindent
which should be understood for $N>2$ of the fact that the $\epsilon^{I}$ are
not linearly independent\footnote{ For $N=2$ we automatically have
  $\mathcal{J}^{I}{}_{J}=\delta^{I}{}_{J}$}.  As a consequence of the above
identity, the contraction of $\mathcal{J}$ with any of the bilinears is the
identity.  Using this result and Eq.~(\ref{eq:mp1}), we find

\begin{equation}
\label{eq:p2}
\Phi^{KL}{}_{ab}=\frac{2 M^{IK}M_{IJ}}{|M|^{2}}\Phi^{JL}{}_{ab}=  
\frac{2 M^{IK}}{|M|^{2}}V^{L}{}_{I\, [a}V_{b]} 
-i\frac{M^{IK}}{|M|^{2}}\epsilon_{ab}{}^{cd}V^{L}{}_{I\, c}V_{d}\, .
\end{equation}

Other useful identities are

\begin{equation}
\label{eq:MMJJ}
\frac{M_{IJ}M^{KL}}{|M|^{2}} =
\mathcal{J}^{K}{}_{[I}\mathcal{J}^{L}{}_{J]}\, ,
\end{equation}

\noindent
and 

\begin{equation}
\mathcal{J}^{I}{}_{J}=\delta^{I}{}_{J}-\tilde{\mathcal{J}}^{I}{}_{J}\, ,  
\end{equation}

\noindent
where

\begin{equation}
\label{eq:complementaryprojector}
\tilde{\mathcal{J}}^{I}{}_{J} \equiv 
 \frac{(N-2) \tilde{M}^{IK_{1}\cdots K_{N-3}}
 \tilde{M}_{JK_{1}\cdots K_{N-3}}}{|\tilde{M}|^{2}}\, ,
\hspace{1cm}
|\tilde{M}|^{2} \equiv \tilde{M}^{I_{1}\cdots I_{N-2}}
 \tilde{M}_{I_{1}\cdots I_{N-2}}
= \frac{(N-2)!}{2}|M|^{2}\, ,
\end{equation}

\noindent
is the complementary projector.

We can always use the 1-form $\hat{V}\equiv V_{\mu}dx^{\mu}$ to construct the
$0^{th}$ component of a Vielbein basis $\{e^{a}\}$

\begin{equation}
\label{eq:e0}
e^{0}\equiv \tfrac{1}{\sqrt{2}}|M|^{-1}\hat{V}\, .  
\end{equation}

\noindent
Let us define the three 1-forms 

\begin{equation}
\label{eq:1formsVm}
\hat{V}^{m} \equiv |M| e^{m}\, ,
\hspace{1cm}
m=1,2,3\, ,  
\hspace{1cm}
V^{m\, \mu}V^{n}{}_{\mu}= -|M|^{2}\delta^{mn}\, ,
\end{equation}

\noindent
and the  spacetime-dependent Hermitean matrices 

\begin{equation}
\label{eq:sigmamdef}
(\sigma^{m})^{I}{}_{J}\equiv -\sqrt{2}\, V^{m\, \mu}V^{I}{}_{J\, \mu}\, ,  
\end{equation}

\noindent
so we can decompose the 1-forms $\hat{V}^{I}{}_{J}=V^{I}{}_{J\, \mu}dx^{\mu}$ as

\begin{equation}
\label{eq:decompositionVIJ}
\hat{V}^{I}{}_{J} = \tfrac{1}{2}\mathcal{J}^{I}{}_{J}\hat{V}
+\tfrac{1}{\sqrt{2}}(\sigma^{m})^{I}{}_{J}\hat{V}^{m}\, ,
\end{equation}

\noindent
and

\begin{equation}
\hat{V}^{I}{}_{J\, a}
=\tfrac{1}{\sqrt{2}}|M|\left[\delta_{a}{}^{0}\mathcal{J}^{I}{}_{J}
+\delta_{a}{}^{m}(\mathcal{\sigma}^{m})^{I}{}_{J} \right]\, .  
\end{equation}

\noindent
While this decomposition is unique, the matrices $\sigma^{m}$ are defined only
up to local $\mathrm{SO}(3)$ rotations of the $\hat{V}^{m}$.

The properties satisfied by the 1-forms $\hat{V}^{I}{}_{J}$ can be used to
prove the following properties for the $\sigma^{x}$ matrices:

\begin{eqnarray}
\label{eq:sigmaprop1}
\sigma^{m}\sigma^{n} & = & \delta^{mn}\mathcal{J} +i
\varepsilon^{mnp}\sigma^{p}\, ,\\
& & \nonumber \\
\label{eq:sigmaprop2}
\mathcal{J}\sigma^{m} &  = &  \sigma^{m} \mathcal{J} = \sigma^{m}\, ,\\
& & \nonumber \\
\label{eq:sigmaprop3}
(\sigma^{m})^{I}{}_{I} & = & 0\, ,\\  
& & \nonumber \\
\label{eq:sigmaprop4}
\mathcal{J}^{K}{}_{J}\mathcal{J}^{L}{}_{I}
& = & 
\tfrac{1}{2}\mathcal{J}^{K}{}_{I}\mathcal{J}^{L}{}_{J}
+
\tfrac{1}{2}(\sigma^{m})^{K}{}_{I}(\sigma^{m})^{L}{}_{J}\, ,\\
& & \nonumber \\
\label{eq:sigmaprop5}
M_{K[I}(\sigma^{m})^{K}{}_{J]} & = & 0\, ,\\  
& & \nonumber \\
\label{eq:sigmaprop6}
2|M|^{-2}M_{LI}(\sigma^{m})^{I}{}_{J}M^{JK} & = & (\sigma^{m})^{K}{}_{L}\, ,\\  
& & \nonumber \\
\label{eq:sigmaprop7}
|M|^{-2}M^{IJ}M_{KL} & = &
-\tfrac{1}{3}(\sigma^{m})^{[I}{}_{[K}(\sigma^{m})^{J]}{}_{L]}\, ,\\
& & \nonumber \\
\label{eq:sigmaprop8}
(\sigma^{[m|})^{I}{}_{J}(\sigma^{|n]})^{K}{}_{L}
& = & 
-\tfrac{i}{2}\varepsilon^{mnp}[\mathcal{J}^{I}{}_{L}(\sigma^{p})^{K}{}_{J}
-(\sigma^{p})^{I}{}_{L}\mathcal{J}^{K}{}_{J}]\, .
\end{eqnarray}

\noindent
That is: they, together with $\mathcal{J}$, generate a $\mathfrak{u}(2)$
subalgebra of $\mathfrak{u}(N)$ in the eigenspace of $\mathcal{J}$ of
eigenvalue $+1$ and provide a basis in the space of Hermitean matrices
satisfying $\mathcal{J}A\mathcal{J}=A$: the last of the above properties
is a completeness relation in that subspace since it implies that 

\begin{equation}
A^{L}{}_{J}  
=
\mathcal{J}^{L}{}_{I} A^{I}{}_{K}  \mathcal{J}^{K}{}_{J}
= 
\tfrac{1}{2}\mathrm{Tr} \left(A\mathcal{J}\right) \mathcal{J}^{L}{}_{J}
+
\tfrac{1}{\sqrt{2}}
\mathrm{Tr}\left[ \tfrac{1}{\sqrt{2}}A\sigma^{m}\right]
(\sigma^{m})^{L}{}_{J}\, .
\end{equation}

\noindent
Then, if $A$ is an
$N\times N$ Hermitean matrix such that $\mathrm{Tr}\, 
(A\mathcal{J})=\mathrm{Tr}\, (A\sigma^{x})=0\, ,\,\,\,\forall_{x=1,2,3}$, it
satisfies $\mathcal{J}A\mathcal{J}=0$ and it can be written in the form

\begin{equation}
A = (1-\mathcal{J}) A\mathcal{J} +\mathcal{J}A(1-\mathcal{J})
+(1-\mathcal{J}) A(1-\mathcal{J})\, .
\end{equation}

It is not clear when a combination of global $\mathrm{U}(N)$ and local
$\mathrm{SO}(3)$ transformations is enough to render the matrices $\sigma^{x}$
constant; however, whenever it is possible, then the projector $\mathcal{J}$
will also be constant.  Needless to say, in the $N=2$ case it is always
possible.


\section{Connection and curvature of the conforma-stationary metric}
\label{sec-conformastationarymetric}

A conforma-stationary metric has the general form

\begin{equation}
ds^{2} = |M|^{2}(dt+\omega)^{2} 
-|M|^{-2}\gamma_{\underline{m}\underline{n}}dx^{m}dx^{n}\, ,
\hspace{1cm}
m,n=1,2,3\, ,
\end{equation}

\noindent
where all components of the metric are independent of the time coordinate $t$.
Choosing the Vielbein basis

\begin{equation}
(e^{a}{}_{\mu}) = 
\left(
  \begin{array}{cc}
|M| & |M| \omega_{\underline{m}} \\
& \\
0 & |M|^{-1} v_{\underline{m}}{}^{n} \\
  \end{array}
\right)\, ,
\hspace{1cm}
(e^{\mu}{}_{a}) = 
\left(
  \begin{array}{cc}
|M|^{-1} & -|M| \omega_{m} \\
& \\
0 & |M| v_{m}{}^{\underline{n}} \\
  \end{array}
\right)\, ,
\end{equation}

\noindent
where 

\begin{equation}
\gamma_{\underline{m}\underline{n}}
=v_{\underline{m}}{}^{p}v_{\underline{n}}{}^{q}\delta_{pq}\, ,
\hspace{1cm}
v_{m}{}^{\underline{p}}v_{\underline{p}}{}^{n}v_{n}\, ,
\hspace{1cm}
\omega_{m}= v_{m}{}^{\underline{n}}\omega_{\underline{n}}\, ,   
\end{equation}

\noindent
we find that the spin connection components are

\begin{equation}
  \begin{array}{rclrcl}
\omega_{00m} & = & -\partial_{m}|M|\, , \hspace{2cm} &
\omega_{0mn} & = &  \frac{1}{2}|M|^{3}f_{mn}\, ,\\
& & & & & \\
\omega_{m0n} & = & \omega_{0mn}\, , &
\omega_{mnp} & = & - |M| \varpi_{mnp} -2 \delta_{m[n}\partial_{p]}|M|\, ,\\
\end{array}
\end{equation}

\noindent
where $\varpi_{m}{}^{np}$ is the 3-dimensional spin connection and 

\begin{equation}
\partial_{m} \equiv v_{m}{}^{\underline{n}}\partial_{\underline{n}}\, ,
\hspace{1cm}
f_{mn}=  v_{m}{}^{\underline{p}}  v_{n}{}^{\underline{q}}
f_{\underline{p}\underline{q}}\, ,  
\hspace{1cm}
f_{\underline{m}\underline{n}} \equiv
2\partial_{[\underline{m}}\omega_{\underline{n}]}\, .
\end{equation}

The components of the Riemann tensor are 

\begin{equation}
  \begin{array}{rcl}
R_{0m0n} & = & {\textstyle\frac{1}{2}}\nabla_{m}\partial_{n}|M|^{2}
+\partial_{m}|M|\partial_{n}|M| -\delta_{mn}(\partial|M|)^{2}
+{\textstyle\frac{1}{4}}\nabla{m}|M|^{6}f_{mp}f_{np}\, , \\
& & \\
R_{0mnp} & = & -{\textstyle\frac{1}{2}}\nabla_{m}(|M|^{4}f_{np})
+{\textstyle\frac{1}{2}}f_{m[n}\partial_{p]}|M|^{4}
-{\textstyle\frac{1}{4}}\delta_{m[n}f_{p]l}\partial_{q}|M|^{4}\, ,\\
& & \\
R_{mnpq} & = & -|M|^{2}R_{mnpq} 
+{\textstyle\frac{1}{2}}|M|^{6}(f_{mn}f_{pq}-f_{p[m}f_{n]q})
-2\delta_{mn,pq}(\partial|M|)^{2}
+4|M|\delta_{[m}{}^{[p}\nabla_{n]}\partial^{q]}|M|\, ,\\
  \end{array}
\end{equation}

\noindent
where all the objects in the right-hand sides of the equations are referred to
the 3-dimensional spatial metric $\gamma$ and the 3-dimensional spin
connection $\varpi$.  The components of the Ricci tensor are

\begin{equation}
  \begin{array}{rcl}
R_{00} & = & -|M|^{2} \nabla^{2}\log{|M|} 
-{\textstyle\frac{1}{4}}|M|^{6}f^{2}\, ,\\
& & \\
R_{0m} & = &  {\textstyle\frac{1}{2}} \nabla_{n}(|M|^{4}f_{nm})\, ,\\
& & \\
R_{mn} & = & |M|^{2}
\{
R_{mn} +2\partial_{m}\log{|M|}\partial_{n}\log{|M|}
-\delta_{mn}\nabla^{2}\log{|M|}
-{\textstyle\frac{1}{2}}|M|^{4}f_{mp}f_{np}
\}\, ,\\
  \end{array}
\end{equation}

\noindent
and the Ricci scalar is

\begin{equation}
R=-|M|^{2}
\{
R -{\textstyle\frac{1}{4}}|M|^{4}f^{2} -2 \nabla^{2}\log{|M|}
+2(\partial\log{|M|})^{2}
\}\, ,  
\end{equation}


%
\end{document}